\documentclass[12pt]{article}
\usepackage{amsmath}
\usepackage{mathtools}
\usepackage{graphicx}
\usepackage[utf8]{inputenc}
\begin{document}
\begin{center}
\textbf{On Models of String Cosmology}\\
\vspace{5mm}
 Poula Tadros\footnote{Email: pttadr@utu.fi} and Moahmed Assad Abdel-Raouf  \footnote{Email: assaad@sci.asu.edu.eg}\\

 \vspace{5mm}
$^{[1]}$Physics and Astronomy department, Faculty of Science and Engineering, The University of Turku, Finland.\\
$^{[2]}$Physics department, Faculty of Science, Ain Shams University, Cairo, Egypt.\\
\end{center}
\vspace{10mm}
\begin{abstract}
 In this work the most promising string cosmology models are reviewed model by model in detail. These models are based on concepts form string theory and heterotic M theory to construct possible scenarios for the beginning and the evolution of the universe. In all models as in standard cosmology the problem of particle-antiparticle asymmetry arises, the issue is that all models predict equal number of matter and antimatter to be present in the universe, this contradicts the observations that the number of antiparticle is much less than the number of particles. In order to solve the asymmetry dilemma of particles and antiparticles, it is proposed that particles and antiparticles were distributed in the brane and antibrane, respectively, immediately after the collision. Implications of these scenarios are discussed in detail.
\end{abstract}

\textbf{Declarations}\\
The authors have no relevant financial or non-financial interests to disclose.\\
\newpage

\tableofcontents
 \section{\textbf{Introduction}}

In this work we present a brief account on string cosmology. Our main aim is to simplify the terminology and summarize the fundamental aspects of the models. The review is divided into three parts: the first is a review on standard cosmology from Copernicus principle and first assumptions to the different types of inflation which are required for string cosmology. The second part is devoted to the presentation of certain aspects of string theory used directly in the main subject including D-branes, Horava-Witten theory and string thermodynamics. The last part of the review we give a comprehensive discussion on string cosmology models highlighting each model's achievements and problems to be solved.\\

\section{\textbf{Part 1 : Cosmology}} 
Cosmology is the study of the universe at a large scale, linguistically cosmos means world and logos means study or knowledge, so cosmology literally means the study of the world.\\   
In this part the major aspects of classical cosmology is briefly reviewed as an introduction to part 3 "brane world cosmologies".\\

\subsection{\textbf{Hubble's law}}
The first postulate of cosmology is that the universe is isotropic i.e. it looks the same regardless of the direction of observation which means it possesses spherical symmetry. The second postulate is given by Copernicus cosmological principle which states that we, as observers on earth, are not special in the universe i.e. every point in the universe is the same as the point we are observing from.\\
 As a direct consequence it is concluded that the universe must look isotropic at every point so is homogeneous i.e. possesses translational symmetry where homogeneity and isotropy are defined in a mathematically rigorous way in Ref$[1]$. It is obvious that this is wrong on small scales (up to a galactic scale), for example the sun,which is much more massive than all the planets so our solar system, is locater in the certer so the solar system is not homogeneous nor isotropic from the earth's point of view, but averaged over large scales of 100 megaparsecs(Mpc) it is observed to be true(Kirshner et al. 1981)$[2]$. \\
To build up our physics we choose our coordinates to be a three dimensional grid or lattice with galaxies are fixed on the lattice points. (here we approximate each galaxy as point since we are considering a much larger scale). Let the coordinate distance between lattice points to be $\textbf{X}=(x,y,z)$
as in figure 1 below.\\
\begin{figure}[hbtp]
\centering
\includegraphics[scale=0.25]{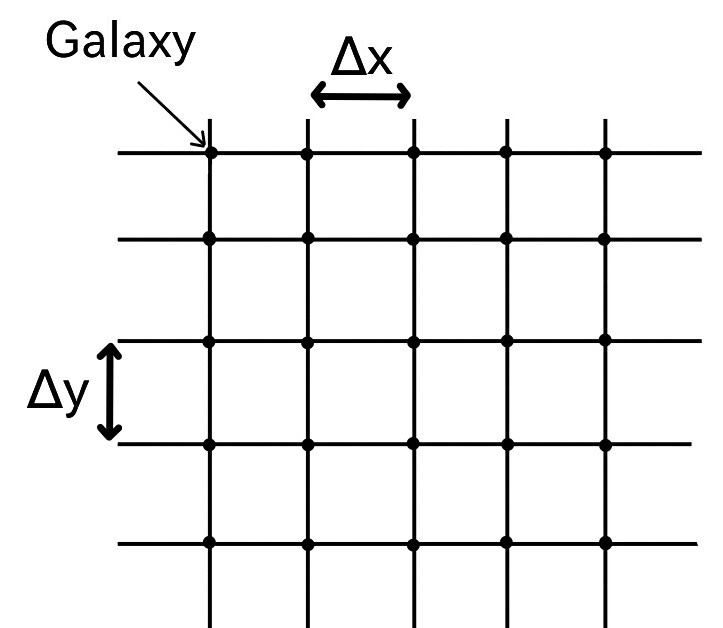}
\caption{a two dimensional lattice structure of galaxies on a large scale, with a galaxy on each lattice point}
\end{figure}

One may think that the lattice structure will be destroyed by the motion of the galaxies with respect to each others , but observations tells us that galaxies move coherently -up to a peculiar motion- and preserve the lattice structure.\\
The physical distance $D_{st}$ between the galaxy s and the galaxy t is proportional to the coordinate distance,the reason of this proportionality is that the coordinate distance is a representation of the physical distance, and the lattice is a representation of the universe on this scale consequently, there is a geometric similarity between the lattice and the universe, and from the geometric similarity the proportionality holds. Because the universe is homogeneous the proportionality factor, which is called the scale factor (a), is independent of the coordinate distance, so it depends only on time i.e. $a=a(t)$.\\
The expression for the physical distance is given by
\begin{equation}
D_{st}=a(t)|\textbf{X}_{st}| \\
\end{equation}
where $\textbf{X}_{st}$ is the coordinate distance between the galaxy s and the galaxy t.\\
To get the velocity the physical distance is differentiated with respect to time,and dropping the suffices because this is a general rule for any two galaxies
\begin{equation}
v = \frac{dD}{dt} = a(t)\frac{dX}{dt}+\frac{da}{dt}X\\
\end{equation}
where $X=|\textbf{X}|$. \\
The first term of the right hand side of eq. (2) is called the peculiar velocity ($v_{pec}$) which represents the motion of galaxies with respect to a comoving observer (the observer with respect to which the lattice points are at rest). Thus it represents dislocations in the lattice structure. The second term of eq.(2) is called the Hubble flow and it represents the velocity due to the stretching or shrinking of the lattice itself i.e. the expansion or contraction of the universe.\\
Thus eq.(1) reduces to 
\begin{equation}
v=v_{pec}+\dot{a}X \\ 
\end{equation}

where $\dot{a}=\frac{da}{dt}$.\\
Now in order to have a coordinate free description of physics, so we should cancel the coordinate distance X using eq(1) and
substituting $X=\frac{D}{a(t)}$ we get
\begin{equation}
v=v_{pec}+\frac{\dot{a}}{a} D
\end{equation}
the factor $\frac{\dot{a}}{a}$ is called Hubble´s parameter and is denoted by H. Also, the peculiar velocity declines as $\frac{1}{a(t)}$, so this term will decrease until the velocity converges to the Hubble flow
\begin{equation}
v=HD
\end{equation}
Eq. (5) is called the Hubble law.\\
This means that wherever the observer is located every galaxy is receding away from him,and leads to the conclusion that we are on an expanding surface i.e. the universe is expanding.\\

\subsection{\textbf{Friedmann Robertson Walker (FRW) spacetime}}
The FRW metric is a solution of Einstein's field equations,It was worked out by Friedmann$[3]$, and independently by Robertson$[4]$ and Walker$[5]$ by assuming homogeneity and isotropy only.\\
Homogeneity and isotropy of space require that the space has a constant curvature (K), and by a simple rescaling one can set $K=0,1,-1$ for a flat,positively curved and negatively curved space respectively.\\
The FRW metric is then given by
\begin{equation}
ds^2=-dt^2+a^2(t)(\frac{dr^2}{1-Kr^2}+r^2d\Omega^2)
\end{equation}
where $d\Omega^2=d\theta^2+sin^2\theta d\phi^2$ and $r,\theta , \phi$ are the usual spherical coordinates.\\
In our case it is not the most natural coordinate to use because we are living on the surface not in the ambient higher dimensional space as in figure 2.\\ Defining a new coordinate $\chi$ by
\begin{equation}
r=S_K(\chi)= 
\left \{ 
\begin{array}{cc}
sin \chi & K=1 \\ 
\chi & K=0 \\ 
sinh \chi & K=-1
\end{array}
\right .
\end{equation} 

\begin{figure}[hbtp]
\centering
\includegraphics[scale=1]{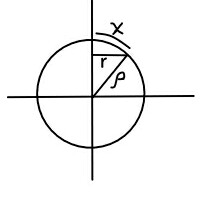}
\caption{with $\rho =1$ as set by rescaling, $\chi$ is the arc length from an observer sitting in the north pole of the circle}
\end{figure}

Substituting in eq.(6) we get
\begin{equation}
ds^2=-dt^2+a^2(t)[d\chi^2+S^2_K(\chi)d\Omega^2]
\end{equation}
This is the most natural metric used as shown in figure 2
if we want to treat time and space equally, we define the conformal time $\eta$ by 
\begin{equation}
 dt=a(t)d\eta \rightarrow \eta-\eta_0=\int_{t_0}^t\frac{dt'}{a(t')}
\end{equation}
it is called conformal because it is the usual time but rescaled by the scale factor.\\
In this case the metric will be
\begin{equation}
ds^2=a^2(t)[-d\eta+d\chi^2 + S^2_K(\chi) d\Omega^2]
\end{equation}
After deriving the metric in various coordinates, the dynamics in the FRW spacetime is studied, and the equations of motion is derived by Friedmann using the most general stress-energy tensor consistent with the homogeneity and isotropy(the perfect fluid) leads to
\begin{equation}
T_{\mu\nu}=Pg_{\mu\nu}+(P+\rho)u_{\mu}u_{\nu}
\end{equation}
where P is the pressure made by the matter content and $\rho$ is the energy density and $g_{\mu\nu}$ is the metric tensor. Substituting in Einstein's field equations we get Friedmann's equations(the detailed calculations are given in Ref[6])
\begin{equation}
H^2=\frac{8 \pi G}{3} \rho-\frac{K}{a^2}
\end{equation}
\begin{equation}
\dot{\rho}=-3H(\rho+P)
\end{equation}
where G is Newton's constant.\\
Defining the critical density by $\rho_{crit}(t)=\frac{3H(t)^2}{8 \pi G}$, the first Friedmann equation can be written as
\begin{equation}
\frac{K}{a^2}=\frac{8 \pi G}{3}(\rho-\rho_{crit})
\end{equation}
This gives a test for the curvature as $\rho=\rho_{crit}$ then $K=0$ (flat universe),\\ if $\rho>\rho_{crit}$ then $K=1$ (closed (spherical) universe),\\ if $\rho<\rho_{crit}$ then $K=-1$(open(hyperbolic)universe).\\
\subsection{\textbf{Some problems in FRW theory}}
\textbf{1-The flatness problem.}\\
This problem was first stated in Ref[7] and Ref[8]. The cosmological observations tells us that the energy density of the universe is very close to the critical density $\rho=\rho_{crit}(1 \pm 0.01 )$. In the past the energy density was even closer to the critical density, this flavors the choice of $K=0$ (the flat universe) although the observations does not.\\
The problem is why is the energy density is close to the critical density for such a long time?\\
The easiest solution is declaring that the universe is spatially flat, but this will not be a sufficient solution because it is not confirmed by experiments due to experimental errors.\\
\textbf{2- The horizon problem.}\\
This problem was stated in Ref[9].\\
The question is: Why is the universe isotropic? \\
To explain the problem consider a radially moving photon in the FRW spacetime so $d\Omega^2=0$ because it is moving radially and $ds^2=0$ as light is moving on null geodesics. Substituting in eq.(8)
we get
\begin{equation}
0=-dt^2+a^2(t)d\chi^2 
\end{equation}
$$\rightarrow  d\chi^2=\frac{dt^2}{a^2(t)}$$
$$\rightarrow d\chi=-\frac{dt}{a(t)}$$
and the negative sign is because the photons are propagating radially inwards so the radial distance decreases with time.\\
Let $t_0$ be the time of observation(today) and $t_{cmb}$ is the time of release of the cosmic microwave background(CMB), calculating the distance between two opposite points of the CMB which can be observed(the distance traveled by light from the big bang at $t=0$ to $t_{cmb}$) by integrating $d\chi$ and rescaling by the scale factor at $t_0$ to "translate" the distance obtained then to today's distance and taking the absolute value, we get
\begin{equation}
\chi_H=a(t_0)\int_{0}^{t_{cmb}}\frac{dt}{a(t)}
\end{equation}
where $\chi_H=\chi_0$ is the distance between two opposite points of the CMB as observed today.\\
The dominant era then is the matter dominated era, so we use $a(t) \propto t^{2/3}$. Substituting in the integration we get
\begin{equation}
\chi_0=3t_{cmb}^{1/3}t_0^{2/3}
\end{equation}
The distance traveled by light from the CMB release to today is
\begin{equation}
\chi_H=a(t_0)\int_{t_{cmb}}^{t_0}\frac{dt}{a(t)}
\end{equation}
this is the horizon distance (the distance between the farthest two causally connected points). Using the same approximation by considering the scale factor of the matter dominated era, we get
\begin{equation}
\chi_0=3(t_0^{1/3}-t_{cmb}^{1/3})t_0^{2/3}
\end{equation}
The ratio between the two quantities defined in equations (17) and (19) is
\begin{equation}
\frac{2\chi_0}{\chi_H}=\frac{2(t_0^{1/3}-t_{cmb}^{1/3})}{t_{cmb}^{1/3}}
\end{equation}
The numerical actor 2 appearing in the left hand side of equation (20) is because we are considering two opposite points i.e. in fact two horizons.\\
Setting $t_0>>t_{cmb}$ neglecting $t_{cmb}^{1/3}$ in the numerator we get
\begin{equation}
\frac{2\chi_0}{\chi_H}=2(\frac{t_0}{t_{cmb}})^{1/3}
\end{equation}
putting the numbers we find that the ratio is about 72.\\
This means that the distance between two opposite points in the CMB is much larger than the distance of the causally connected region of the two points, the only conclusion is that the two points are not causally connected as demonstrated in figure 3, and this is the horizon problem, why are they looking the same and in thermal equilibrium although they were never causally connected?\\
\begin{figure}[hbtp]
\centering
\includegraphics[scale=0.25]{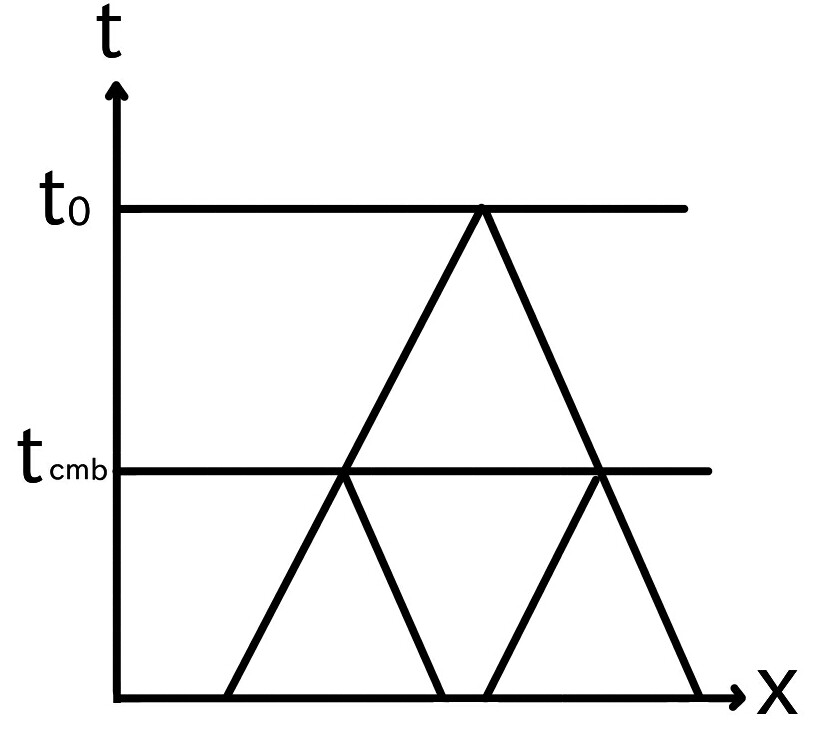}
\caption{Today at $t=t_0$ the CMB regions observed at opposite sides are influenced by the region bounded by light rays trajectories at the time the CMB is formed $t_{CMB}$,these light rays can only be influenced by regions bounded by light rays trajectories starting at the big bang. These regions are not causally connected, so why the light rays are so similar?}
\end{figure}

\textbf{3-Unwanted relics problem.}\\

Why is the relics density so small in the universe?\\
The term Relics refers to some structures produced during phase transitions, like the great unified theory (GUT) phase transition, such as magnetic monopoles, cosmic strings or domain walls or any other topological defect.\\
let us take monopoles as an example, Kibble mechanism$[10,11]$ generates one monopole per causally connected horizon, and as the temperature drops through the phase transition it is expected to create one monopole per nucleon. On the other hand observations tell us that monopoles (and relics in general) is much smaller in fact no monopole or cosmic string was observed yet$[12]$.\\
\subsection{\textbf{Inflation}}
Inflation is a postulate or a framework was first proposed by Starobinski$[13]$ and then by Alan Guth$[6]$ to solve the problems of standard cosmology, and then many inflationary models were proposed later.\\
The general idea is that the universe underwent a period of exponential expansion before the radiation dominated era. Various models were proposed to describe this inflationary phase, as in quantum field theory each model is characterized by an action.\\
At first define the comoving Hubble's radius as
\begin{equation}
r=\frac{1}{aH}=\frac{1}{\dot{a}}
\end{equation}
It represents the radius of the causally connected patch with respect to a point at a given time i.e. $\chi>r$ means particles cannot transfer signals to each other now but maybe in the past they could or in the future they will can communicate. Differentiating r with respect to time leads to
$$\dot{r}=\frac{-\ddot{a}}{\dot{a}^2}$$
this means that $\dot{r}<0$ if and only if $\ddot{a}>0$
i.e. the expansion of the universe is accelerating (the universe is inflating) if and only if the comoving Hubble's radius is decreasing with time.\\
This framework solves the three problems discussed earlier as during an exponential inflation the comoving Hubble's radius decreases rapidly i.e. the universe began with some comoving Hubble's radius and then it decreases until a much smaller value until the radiation dominated era then it is slowly increasing again.\\
- The flatness problem is automatically solved by stating that we are observing a very small patch of the true universe so we can not observe its curvature (we can observe only distances less than or equal the comoving Hubble's radius which is very small now).\\
- The horizon problem is solved because before inflation the comoving Hubble's radius was so large that the areas which are seen by us as causally disconnected now due to the small comoving Hubble's radius now were in fact connected in the past (figure 4). This explains why we see isotropic universe and that the CMB in opposite directions is in thermal equilibrium.\\
\begin{figure}[hbtp]
\centering
\includegraphics[scale=0.25]{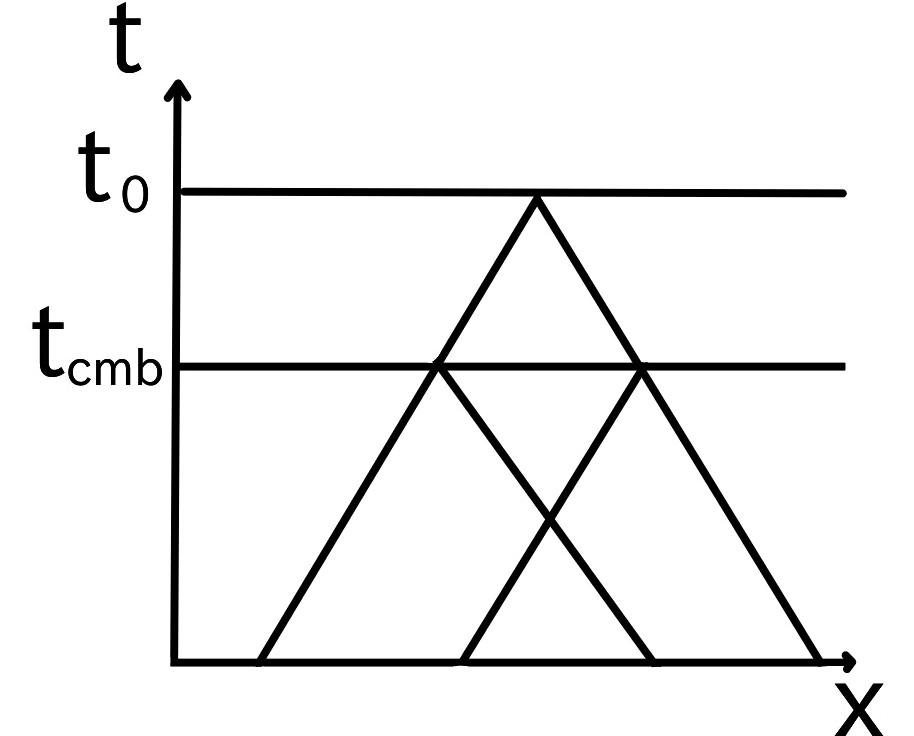}
\caption{Inflation solves the horizon problem by buying time or equivalently a larger connected horizons so that the seemingly disconnected parts as seen today were actually connected and in thermal equilibrium before inflation}
\end{figure}

- The unwanted relics problem is solved similar to the horizon problem because in the past the number of disconnected regions is small so the number of relics(for example monopoles) is as well small.\\
In any inflationary model two conditions must be satisfied, the first is that the comoving Hubble's radius is decreasing rapidly 
\begin{equation}
\frac{dr}{dt}=-\frac{\dot{a}(t)H+a(t)\dot{H}}{a(t)^2H^2}
= -\frac{\dot{a(t)}H+a(t)\dot{H}}{\dot{a(t)}^2}
\end{equation}
multiplying and dividing by a(t) and using eq. (22)
$$\frac{-1}{a(t)}[\frac{\dot{a}(t)a(t)H+a^2(t)\dot{H}}{\dot{a}^2(t)}]=\frac{-1}{a(t)}[\frac{\dot{a}(t)^2+a(t)^2\dot{H}}{\dot{a}(t)^2}]=\frac{-1}{a(t)}[1+\frac{\dot{H}}{H^2}]=\frac{-1}{a(t)}[1-\epsilon]$$ 
where $\epsilon=\frac{-\dot{H}}{H^2}$
the condition is that $\dot{r}<<0$ or $\epsilon<<1$\\
The second condition is that the inflation lasts sufficiently long to solve the problems (at least 60 e-folds)
\begin{equation}
|\eta|=|\frac{\dot{\epsilon}}{H\epsilon}|<<1
\end{equation}
i.e. the fractional change of $\epsilon$ per Hubble's time is small.\\
Now we derive the condition the equation of state of the fluid must satisfy to drive an inflation.\\
Beginning with the Friedmann equations and neglecting the curvature contribution, we get 
\begin{equation}
H^2=\frac{8\pi G}{3}\rho \ \ , \ \ \dot{\rho}=-3H(\rho+P)
\end{equation}
Differentiating the first equation with respect to time
\begin{equation}
2H\dot{H}=\frac{8 \pi G}{3}\dot{\rho}
\end{equation}
substituting $\dot{\rho}$ from the eq. (25), yields
$$2H\dot{H}=-\frac{8 \pi G}{3}[3H(\rho+P)]$$  
or $\dot{H}=4 \pi G(\rho+P)$ using $$H=\frac{\dot{a}}{a} \rightarrow \dot{H}=\frac{a\ddot{a}-\dot{a}^2}{a^2}$$
substituting we get
$$
\frac{\ddot{a}}{a}-(\frac{ \dot{a}}{a})^2=-4 \pi G (\rho+P)
$$

We then get
\begin{equation}
\frac{\ddot{a}}{a}-H^2=-4 \pi G(\rho +P)
\end{equation}
by the first Friedmann equation 
$$\frac{\ddot{a}}{a}-\frac{8 \pi G}{3}\rho=-4 \pi G(\rho +P)$$
$$\rightarrow \frac{\ddot{a}}{a}=-\frac{4 \pi G}{3}(\rho +3P)$$
This is called the acceleration equation, In any inflationary model $\ddot{a}>0$ so $\rho+3P<0$ so in a linear equation of state $P=w\rho$, we must have $w<-\frac{1}{3}$.
This violates the strong energy conditions which indicates that this region is unreliable for classical general relativity.\\
Under the previous conditions many inflationary models were proposed.\\
1-the "old" inflation$[14]$.\\
2-The false vacuum inflation$[15]$, where the inflation is caused by the Higgs field being trapped in a false vacuum state.\\
3-The hybrid inflation$[16]$, where there are two scalar fields one to drive inflation and the other to stop it.\\
4-Kinetically driven inflation(K inflation) $[17]$,Dirac-Born-Infeld (DBI)infaltion is an important special case, where higher order kinetic terms of a field are the cause of inflation.\\
5-Gauge field driven inflation$[18]$,where the inflation is driven by a non-Abelian gauge field, this type of inflation satisfies the slow roll conditions.\\
6-Power law inflation$[19]$,where the scale factor is proportional to a power of time $a(t) \propto t^p$ where p is a positive number.\\
7-Chaotic inflation$[20]$ where many patches of the spacetime manifold underwent inflation, one of which is our universe.\\
8-Dante's Inferno$[21]$ where a high scale inflation happens in a very small region of field space.\\
6-The slow roll inflation$[22,23]$ where inflation is driven by a single scalar field whose kinetic term is much smaller than the potential term until the end of inflation.\\
7-Some recent models combining more than one of the above models like combining DBI and slow roll inflation$[24]$.\\
Slow roll inflation model is the most important one for the purpose of the review so it is discussed in detail. In this model the inflation was driven by a single scalar field called the inflaton $\phi$ with the action
\begin{equation}
  S=\int d^4x\sqrt{g}[\frac{R}{16 \pi G}-\frac{1}{2}g^{\mu \nu}\partial_{\mu}\phi \partial_{\nu}\phi -V(\phi)]
\end{equation}
where $g^{\mu \nu}$ is the metric tensor, R is the Ricci scalar and $g=det(g^{\mu \nu})$. \\
Euler lagrange equations for the scalar field is the same as Klein Gordon equation(because the terms in the action involving the field is the same as Klein Gordon action)
\begin{equation}
\partial_{\mu}\partial^{\mu}\phi - V'(\phi)=0
\end{equation}
where V' is the derivative with respect to $\phi$.\\
The energy momentum tensor is the same as Klein Gordon field and is determined by
\begin{equation}
 T_{\mu\nu}=\partial_{\mu}\phi \partial_{\nu}\phi -\frac{1}{2}g_{\mu\nu}\partial_{\sigma}\phi \partial^{\sigma} \phi -g_{\mu\nu}V(\phi)
\end{equation}
Comparing eq. (30) with the perfect fluid energy momentum tensor 
\begin{equation}
 T_{\mu\nu}=Pg_{\mu\nu}+(P+\rho)u_{\mu}u_{\nu}
 \end{equation}
because as stated above it is the most general energy momentum tensor satisfying homogeneity and isotropy.\\
we get
$$\rho = -\frac{1}{2}\partial_{\sigma}\phi \partial^{\sigma} \phi +V(\phi)$$ 

\begin{equation}
P=-\frac{1}{2}\partial_{\sigma}\phi \partial^{\sigma} \phi -V(\phi) 
\end{equation}

$$u_{\mu}=-\frac{\partial^{\mu}\phi}{\sqrt{-\partial_{\sigma}\phi \partial^{\sigma} \phi}}$$  
then 
\begin{equation}
    w-\frac{P}{\rho}=\frac{-\frac{1}{2}\partial_{\sigma}\phi \partial^{\sigma} \phi -V(\phi)}{ -\frac{1}{2}\partial_{\sigma}\phi \partial^{\sigma} \phi +V(\phi)}
\end{equation}
To get an inflation it is assumed that the field is smooth by stating that the the kinetic term is much less than the potential term. Neglecting V we get $w=-1$ and using second Friedmann equation we get $\rho=constant$.\\
This corresponds to a repulsive gravity or an always repulsive force.\\ 
Applying the two conditions to guess the shape of the field, to be consistent with the FRW background symmetries we require that $\phi $ depends only on time, substituting in Friedmann equations yields
\begin{subequations}
\begin{align}
H^2=\frac{8 \pi G}{3}[\frac{-1}{2}\dot{\phi}^2+V(\phi)] \\
-\dot{\phi}\ddot{\phi}+V'\dot{\phi}=3H\dot{\phi}^2 
\end{align}
\end{subequations}
differentiating the first equation with respect to time

$$2H\dot{H}=\frac{8 \pi G}{3}\dot{\phi}(-\ddot{\phi}+V')$$
 
by the second equation
\begin{equation}
    \dot{H}=4 \pi G \dot{\phi}^2
\end{equation}

The first condition is
\begin{equation}
    \epsilon=-\frac{\dot{H}}{H^2}=\frac{4 \pi G \dot{\phi}^2}{H^2}<<1
\end{equation}

i.e. the kinetic term is much less than the potential as stated, then the slope of the field is small (it is indeed a slow roll).\\
The second condition : 

   $$ |\eta|=|\frac{\dot{\epsilon}}{H\epsilon}|$$
  
but using the first condition  $\dot{\epsilon}=8 \pi G \dot{\phi}[\frac{-H\ddot{\phi}+\dot{H}}{H^3}]$  dividing by $H\epsilon$
$$|\eta|=|8 \pi G \dot{\phi}[\frac{-H\ddot{\phi}+\dot{H}}{H^2 \dot{H}}]|$$
$$=|2(-4 \pi G \dot{\phi} \ddot{\phi}\frac{1}{H \dot{H}}+4 \pi G \frac{\dot{\phi}}{H^2}|)$$
the second term is precisely $\epsilon$, and using $\dot{H}=4 \pi G \dot{\phi}^2$ yields\\
$$|\eta|=2|(\epsilon-\frac{\ddot{\phi}}{H \dot{\phi}})|$$
but $\epsilon<<1$, so the second condition is
\begin{equation}
    |\frac{\ddot{\phi}}{H\dot{\phi}}|<<1
\end{equation}

i.e. the second derivative is even much smaller than the first derivative(a nearly flat field).\\
\subsection{\textbf{Some Open problems}}
1- Matter-antimatter asymmetry: one of the most difficult dilemmas is the fact that the number of anti particles is minimal in comparison to the number of particles.\\
In order to have plausible explanation of this dilemma, some conditions are proposed called the sakharov conditions$[25]$ which are the conditions necessary for the matter-antimatter asymmetry to occur but the origin of this asymmetry is yet unknown.\\
2- The transplanckian problem$[26]$ : the physics in the late universe seems to depend on the physics in the scales at the start of the inflation(the physical wavelength corresponding to the large scale structures were shorter than Planck's length), in this scale it is believed that our theories do not work and there is a unified theory the problem is why we can calculate the fluctuations spectrum and get the right answer although our theories do not work at the beginning of the inflation?\\
3- what is the nature of the dark matter?\\
Many evidences tell us that there is another type of massive matter which fall out of the equilibrium with the other contents of the universe earlier than the usual matter exists but its identity is unknown. There are candidates of dark matter but no proven result came yet.\\
4- what is the cause of the accelerated expansion of the universe? the reason is called the dark energy but it is of unknown nature(whether it is the cosmological constant itself or another field). A review on the candidates for dark matter and dark energy is found in Ref[27].\\
5- The cosmological constant problem$[28]$ : Quantum field theory predicts that the vacuum energy of the universe is huge however, it is observed to be very small(discrepancy of 120 orders of magnitude) the explanation of this huge discrepancy is unknown.\\

\section{\textbf{Part 2 : String theory and M-theory}}
This part is a short review of the concepts and aspects of string theory and M-theory which are crucial for the discussion of the string cosmology. The reasons why string theory existed in the first place are reviewed briefly as well.\\
\subsection{\textbf{The history of string theory}}
In 1968 Veneziano proposed a formula for the scattering amplitudes in hadronic systems interacting via strong interactions, this formula turns out to satisfy Regge behavior and the right asymptotic conditions, motivated and proved to be so in Ref[29], and was generalized later$[30]$.\\
Veneziano formula was understood by Susskind$[31]$, Nambu$[32]$(Reviewed in Ref[33]) to be the formula for the scattering amplitude of fundamental relativistic strings with quarks at its endpoints. This is by the way not so different from our current understanding in the regime of quantum chromodynamics with flux tubes connecting quarks, and the excitations of the strings represent mesons and hadrons . However, this new framework failed to describe strong interactions mainly due to three reasons (beside other technical problems such as its failure to get fit with the patron properties reviewed in Ref[29]): \\
1- The theory makes sense only in 26 dimensions as will be proven later.\\
2- Excitations of fundamental strings proposed by this framework contain spin 2 massless particles and spin 1 particles which have no existence in hadronic world.\\
3- The theory contains only bosons.\\
Another caveat in the theory which is not related to hadrons but is a devastating trait in a theory is the fact that the theory contains tachyons which is not necessary an inconsistency but it is an instability in the vacuum state.\\       
All these lead to the shift of interest form the new theory (which was named string theory and then bosonic string theory afterwards)
to a gauge field theory for strong interactions i.e. quantum chromodynamics(QCD) which is still used until now. However, another model was present by Ramond, Neveau and Schwartz$[34,35]$ who introduced fermions to string theory using world sheet supersymmetry(SUSY),defined later in this part,solving the second problem above, restricting the dimensions to 10, but it was not of any help to the describtion of hadrons Later Wess and Zumino$[36]$(see also later developments presented in Ref[37]) used the idea of spacetime supersymmetry as a generalization of the world sheet supersymmetry in the Ramond-Neveau-Schwartz model. This motivated a lot of work in this direction and the first appearance of supersymmetric gravity(supergravity).\\
In 1977, Gliozzi, Scherk and Olive$[38]$ conjectured that it is possible to modify the Ramond-Neveau-Schwartz model to remove tachyons completely, and the conjecture was proven later in the early 1980s$[39]$. In the same year S-duality was first proposed by Montonen and Olive$[40]$, Two years later it was observed by Osborn$[41]$ that this duality was credible for $\mathcal{N}=4$ super Yang Mills theories, This was evidenced by Sen$[42]$ in 1994 and more by Vafa and Witten$[43]$ who provided even more evidences.\\
In 1984 a paper wrote by John Schwarz and Micheal Green$[44]$ proved with evidences that superstring theory was anomaly free if the gauge group is $E_8$×$E_8$ or $SO(32)$ i.e. consistent and lead does not lead to ultraviolate divergences. In the same year T-duality was proposed by Kikkawa and Yamasaki$[45]$, then in 1986 Strominger$[46]$ published a paper concluding that we can get a large variety of superstring theories by solving the equations for the dilaton and the torsion. the solutions break the gauge groups $E_8$×$E_8$ or $SO(32)$ into many subgroups. This was a disadvantage in the theory because it may give different predictions and will be difficult to know which one (if any) describes our universe.\\
On the other hand, Bernard De Wit,J.Hoppe and H.Nicolai$[47]$ published an article deriving quantum mechanics from 11D supermembrane theory and proved that this can be done only if the membranes are represented as matrices. In 1995 the number of superstring theories has been reduced to five(the five known to us until now) with dualities between them, In the same year E.Witten gave a lecture in a conference in the university of southern California and proposed that these dualities tell us that the five superstring theories are just five manifestations of the same theory which is an 11D theory containing membranes and called it M-theory. Later on J.Polchinski$[48]$ proved that the dualities would never exist unless higher dimensional strings(branes) are present and he called it D-branes(for Dirichlet boundary conditions on it). This shows that an acpect of M-theory is that it must include membranes as proposed by Witten in the conference.\\
In 1996, Strominger and C.Vafa$[49]$ observed that the modified theory with D-branes describe a black hole in the theory. Then, the work of J.Maldacena$[50]$ showed the close relation between the theory and black hole physics, this was the first time string theory entered cosmology. In 1997, Banks,Fisher,Shenker and Susskind$[51]$ proposed that the M-theory proposed by Witten is the same as De Wit's model$[47]$.\\
In the same year, Juan M.Maldacena$[52]$ and In 1998 Witten$[53]$ and Gubser, Klebanov and Polyakov$[54]$ developed a correspondence between gauge theories and gravity which was called ADS/CFT correspondence which kept physicists busy understanding and developing it until now.\\
String theory was criticized in many literature$[55,56]$, string theory lacks any experimental test and has some technical issues yet unsolved, so it is considered as an unproved hypothesis but it is useful to develop mathematical tools for many fields.\\

\subsection{\textbf{Developments in string theory}.}
The main idea of this section is to provide the necessary minimum background required for string cosmology.\\
The detailed calculations and explanation are presented in several string theory books$[57-62]$
\subsubsection{\textbf{Nambu Goto action and Polyakov action.}}
In the case of a point particle the action functional is proportional to the length of the world line swept by the particle during motion, and the equation of motion is given by the trajectory which extremizes the action i.e. extremizes the length of the world line. Similarly, the action of a string is proportional to the area of the world sheet swept by the propagating string, formally, the world sheet is a Riemann surface on which a special type of diffeomorphisms called modular transformations act$[10]$, and the equations of motion are the trajectories extremizing the area of the world sheet.\\ 
As any surface the world sheet is parametrised by two parameters $\sigma_1$ and $\sigma_2$ taking their values from a parameters space and mapped by the coordinate maps to the target space containing the world sheet as a surface, the resulting coordinates on the world sheet are called the string coordinates.\\
\begin{equation}
\vec{X}(\sigma_1,\sigma_2)=(X^0(\sigma_1,\sigma_2),X^2(\sigma_1,\sigma_2),X^3(\sigma_1,\sigma_2),...,X^d(\sigma_1,\sigma_2))
\end{equation}

where d is the space time dimension.\\
The infinitesimal area is the area of the parallelogram whose sides are the infinitesimal velocity vectors corresponding to the infinitesimal rectangle whose sides are $d\sigma_1$ and $d\sigma_2$ mapped by the coordinate map
\begin{equation}
dv_i=\frac{\partial\vec{X}}{\sigma^i}d\sigma^i
\end{equation}
where Einstein summation convention is applied and i=1,2.\\
Computing the infinitesimal area dA
$$dA=|dv_1||dv_2|sin\theta =|dv_1||dv_2|\sqrt{1-cos^2\theta}$$
$$=\sqrt{|dv_1|^2|dv_2|^2-|dv_1|^2|dv_2|^2cos^2\theta}$$
$$=\sqrt{(dv_1.dv_1)(dv_2.dv_2)-(dv_1.dv_2)^2}$$
using the definition of $dv_i$ and integrating over $\sigma_1$ and $\sigma_2$
$$A=\int d\sigma_1 d\sigma_2\sqrt{(\frac{\partial X}{\partial\sigma_1}.\frac{\partial X}{\partial \sigma_1})(\frac{\partial X}{\partial \sigma_2}.\frac{\partial X}{\partial\sigma_2})-(\frac{\partial X}{\partial \sigma_1}.\frac{\partial X}{\partial\sigma_2})^2}$$
in the index notation
\begin{equation}
A=\int d\sigma_1 d\sigma_2 \sqrt{(\frac{\partial X_{\mu}}{\partial \sigma_1}\frac{\partial X^{\mu}}{\partial \sigma_1})(\frac{\partial X^{\nu}}{\partial \sigma_2}\frac{\partial X_{\nu}}{ \partial \sigma_2})-(\frac{\partial X^{\mu}}{\partial \sigma_1}\frac{\partial X^{\mu}}{\partial \sigma_2})^2}.
\end{equation}
The term under the square root is negative so we introduce a minus sign, and define the Nambu-Goto action to be $S=-TA$, where T can be shown by dimensional analysis or by discussion in Ref[11] to be the string tension.
\begin{equation}
S=-T\int d\sigma_1 d\sigma_2 \sqrt{-(\frac{\partial X_{\mu}}{\partial \sigma_1}\frac{\partial X^{\mu}}{\partial \sigma_1})(\frac{\partial X^{\nu}}{\partial \sigma_2}\frac{\partial X_{\nu}}{\partial \sigma_2})+(\frac{\partial X^{\mu}}{\partial \sigma_1}\frac{\partial X^{\mu}}{\partial \sigma_2})^2}
\end{equation}
Since the world sheet is a surface embedded in the d-dimensional space it induces a metric on it defined by the pull-back of the metric of the ambient space, we take the space to be the flat Minkowski space so the induced metric is
\begin{equation}
\gamma_{ab}=\frac{\partial X^{\mu}}{\partial\sigma^a}\frac{\partial X^{\nu}}{\partial\sigma^b}\eta_{\mu\nu}
\end{equation}
this form can be proved easily by the fact that $ds^2$ is parameterization invariant
\begin{equation}
dS^2=\eta_{\mu \nu}dX^{\mu}dX^{nu}=\gamma_{ab}d\sigma^a d\sigma^b
\end{equation}
and the action is
\begin{equation}
S=-T\int d^2\sigma \sqrt{-\gamma}\end{equation}
where $\gamma = det(\gamma_{ab})$ and $d^2 \sigma = d\sigma_1 d\sigma_2$
This action is clearly poincare invariant and reparameterization invariant, but it is rather hard to quantize due to the square root. To treat this inspired by the einbien formulation of the point particle, define a dynamical metric on the world sheet $g_{\mu \nu}$, it is not equal to the induced metric it is a metric of the world sheet as a space itself. The proposed action is the Polyakov action      
\begin{equation}
S_{poly}=-T/2 \int d^2\sigma \sqrt{g}g^{ab}\eta_{\mu \nu}\partial_aX^{\mu}\partial_b X^{\nu}
\end{equation}
where $g=det(g_{ab})$ and $\partial_a = \frac{\partial}{\partial\sigma^a}$ .\\    
Polyakov action possesses the same symmetries as Nambu-Goto action plus the Weyl invariance under which the metric is transformed as 
\begin{equation}
 g_{ab}(\sigma) \rightarrow \Omega^2(\sigma) g_{ab}(\sigma)
\end{equation}
 where $\Omega$ is called the scale factor.\\
 As in the point particle's case imposing the equation of motion of our einbien counterpart(the metric) and eliminating it form the action gives the Nambu-Goto action again, this means that the two actions are equivalent.\\ 
\subsubsection{\textbf{Supersymmetry and Wess-Zumino model.}}
The next ingredient is space time supersymmetry which introduced in string theory by Wess and Zumino$[36]$ to remove tachyons and hence the instability in the ground state in string theory giving the superstring theory.\\
Supersymmetry is a symmetry between bosons and fermions relating bosonic degrees of freedom to fermionic ones. This introduces more ghosts to the theory with central charge (c) 11(defined later) reducing the central charge from 26 to 15, regarding that the theory has equal number of bosonic degrees of freedom (with c= 1), and fermionic degrees of freedom (c=3/2) requiring that the space time dimensions are 10 instead of 26 in bosonic string theory.\\     
Here Wess-Zumino model is reviewed postponing the discussion and the proof of critical dimensions and central charge to a later section.\\
The Wess-Zumino model is the simplest model in which supersymmetry is incorporated, the action is
\begin{equation}
S=\frac{-1}{2}\int d^2x[\partial_{\mu}\phi \partial^{\mu}\phi+\bar{\psi}\gamma^{\mu}\partial_{\mu}\psi],
\end{equation}
where $\bar{\psi}$ is the adjoint spinor of the majorana spinor $\psi$  and $\gamma^{\mu}$ are the Dirac matrices, and $\phi$ is a scalar field.\\
This model contains one scalar field and one spinor, and is invariant under the transformations
$$\delta \phi =\bar{\epsilon}\psi$$
$$\delta \psi = \gamma^{\mu}\partial_{\mu}\phi \epsilon,$$  
  where $\epsilon$ is a parameter.\\
These transformations relates bosonic and fermionic degrees of freedom and since the action is invariant then it is a supersymmetric model. \\  
\subsubsection{\textbf{Born-Infeld electrodymanics.}}
Born-Infeld theory for non linear electrodynamics is used to describe electromagnetic fields on the world volume of D-branes (extended objects in string theory defined later in this part). It reduces to the usual Maxwell theory for small electric and magnetic fields. Also it requires a maximal electric field when the magnetic field is set to zero removing the singularity in Maxwell's theory, Note that the self energy of a point particle in Maxwell's theory is infinite while in Born-Infeld theory it is a finite number.\\
The Lagrangian of the theory is
\begin{equation}
\mathcal{L}=-b^2\sqrt{1-\frac{E^2-B^2}{b^2}-\frac{(E-B)^2}{b^4}}+b^2,
\end{equation}
where E is the electric field, B is the magnetic field and b is a constant which will be proved to be the maximal electric field on the D-brane's world volume for vanishing magnetic field.\\ 
For $B=0$ and $E<<b$, the Lagrangian reduces to
\begin{equation}
\mathcal{L}=-b^2\sqrt{1-(\frac{E}{b})^2-(\frac{E}{b})^4}+b^2
\end{equation}
neglecting the quartic term as $E/b<<1$ and expanding the square root to the first order in E/b we get
\begin{equation}
\mathcal{L}=-b^2(1-\frac{E^2}{2b^2})+b^2=E^2/2=\mathcal{L}_{Maxwell,B=0}
\end{equation}
It is clear from the Lagrangian that the maximal electric field is b, or else the square root will not be a real function of E and B.\\
Born-Infeld Lagrangian can be written in a covariant way to manifest Lorentz invariance as
\begin{equation}
\mathcal{L}=-b^2\sqrt{-det(\eta_{\mu\nu}+\frac{1}{b}F_{\mu\nu})}+b^2
\end{equation}
where $F_{\mu\nu}$ is the field strength tensor.\\
\subsubsection{\textbf{Mode expansion and quantization}}
Mode expansion refers to the solution of the equations of motion derived form Nambu-Goto action, it is called so because the solution is written as an infinite sum of vibrational modes plus a center of mass term with each mode corresponding to a particle, this indicates that one string in string theory represents an infinite tower of particles in the ordinary quantum field theory. For this reason we do not see creation or annihilation operators for strings in string theory but operators raising or lowering the mode of vibration of the same string. It is worth mentioning that there are two types of strings corresponding to the two possible topologies can be defined on a compact one dimensional manifold, one is homeomorphic to an interval on the real line and is called an open string because it has endpoints (the interval's endpoints), or homeomorphic to a circle and is called closed strings.\\
The equations of motion and their solutions are presented in a simple way in Ref[11] and Ref[12], writing the solutions directly here 
\begin{equation}
X^{\mu}=X^{\mu}_R(\tau-\sigma)+X^{\mu}_L(\tau -\sigma)
\end{equation}
with
\begin{subequations}
\begin{align}
X^{\mu}_R(\tau-\sigma)=\frac{1}{2}x^{\mu}+\frac{1}{2}l^2(\tau-\sigma)p^2+\frac{i}{2}l \sum_{n \neq 0}\frac{1}{n}\alpha^{\mu}_n e^{-2in(\tau-\sigma)}\\
X^{\mu}_L(\tau+\sigma)=\frac{1}{2}x^{\mu}+\frac{1}{2}l^2(\tau+\sigma)p^2+\frac{i}{2}l \sum_{n \neq 0}\frac{1}{n}\bar{\alpha}^{\mu}_n e^{-2in(\tau+\sigma)}e^{-2in\tau},
\end{align}
\end{subequations}

For closed strings,
where$X^{\mu}$ are the string coordinates and $X^{\mu}_R(\tau-\sigma)$ represent the right moving strings, $X^{\mu}_L(\tau -\sigma)$ are the left moving strings, $\tau$ is the first parameter parameterizing the world sheet, $\sigma$ is the second parameter and $p^{\mu}$ is a constant vector representing the total momentum of the string.\\
Therefore we have
\begin{equation}
X^{\mu}=x^{\mu}+\frac{1}{2}\tau l^2 p^{\mu}+\frac{i}{2}l \sum_{n \neq 0}\frac{1}{n}(\alpha^{\mu}_n e^{2in\sigma}+\bar{\alpha}^{\mu}_n e^{-2in\sigma})
 \end{equation}   

It is clear that it is periodic with respect to $\sigma$ as it is expected form a closed string.\\
For open strings we have two possible boundary conditions,namely, Dirichlet boundary conditions and Newman boundary conditions. In the first case the mode expansion is

\begin{equation}
X^{\mu}=x^{\mu}_0+\frac{\sigma}{\pi}(x^{\mu}_{\pi}-x^{\mu}_0)+\sum_{n \neq 0}\frac{1}{n}\alpha^{\mu}_ne^{-in\tau}sin(m\sigma)
\end{equation}
where $x^{\mu}_0=X(\tau , \sigma=0)$ and $x^{\mu}_{\pi}=X(\tau , \sigma=\pi)$.\\
In the second case, we have
\begin{equation}
X^{\mu}=x^{\mu}+l\tau p^{\mu}+il\sum_{n \neq 0}\frac{1}{n}\alpha^{\mu}_ne^{-in\tau}cos(m\sigma)
\end{equation}
in all cases $\alpha_n$ are the mode expansion coefficients which will be promoted to operators on quantizing the theory.\\
Before quantizing the theory, let us define the canonical momentum by
\begin{equation}
P^{\mu}=\frac{\partial L_{NG}} {\partial \frac{\partial X^{\mu}}{\partial \tau}}
\end{equation}
where $L_{NG}$ is the Nambu-Goto Lagrangian, then quantize by imposing the equal $\tau$ commutation relations and promote $\alpha$ to be operators
\begin{equation}
[P^{\mu}(\tau , \sigma),X^{\nu}(\tau , \sigma')]=\eta^{\mu \nu} \delta(\sigma-\sigma').
\end{equation}
From that, the commutation relation for the operators $\alpha_n$ are
\begin{equation}
[\alpha_m^{\mu}, \alpha_n^{\nu}]=[\bar{\alpha}_m^{\mu}, \bar{\alpha}_n^{\nu}]=im\eta^{\mu \nu} \delta_{m+n , 0}
\end{equation}
and
$$[x^{\mu},p^{\nu}]=i\eta^{\mu \nu}$$
\subsubsection{\textbf{Constraints on the motion.}}
Beginning with Polyakov action, varying with respect to the metric $g^{\mu \nu}$ will give the stress energy tensor
\begin{equation}
T_{\mu \nu}=\frac{-2}{T\sqrt{-g}}\frac{\partial S}{\partial g^{\mu \nu}}.
\end{equation}
then setting $g^{\mu \nu}=\eta^{\mu \nu}$ the equations of motion of $g^{\mu \nu}$ using the parameterization in Ref[11] reads $T_{\mu \nu}=0$.\\
Using the light cone coordinates $\sigma^{\pm}=\tau \pm \sigma$ the constraints are $(\frac{\partial X}{\partial \sigma^+})^2=(\frac{\partial X}{\partial \sigma^-})^2=0$
for closed strings (open strings are the same by analogy) the conditions are
\begin{subequations}
\begin{align}
\sum_n L_n e^{-in\sigma^-}=0 \\
\sum_n \bar{L}_n e^{-in\sigma^+}=0
\end{align}
\end{subequations}

where 
\begin{subequations}
\begin{align}
L_n=\frac{1}{2}\sum_m \alpha_{n-m} \alpha_m\\
\bar{L}_n=\frac{1}{2}\sum_m \bar{\alpha}_{n-m} \bar{\alpha}_m
\end{align}
\end{subequations}
where $\alpha_0^{\mu}=lp^{\mu}$
any classical solution then must satisfy the condition $L_n=\bar{L}_n=0 \ \  \forall n \in N$, these are the generators of the Witt algebra defined below.\\
To discuss quantum mechanical solutions we need to define the Fock space and to introduce ghosts.\\
\subsubsection{\textbf{Fock space and Ghosts.}}
Observing that the equations of motion are wave equations, the Fock space is defined as of a harmonic oscillator, define annihilation and creation operators by
\begin{equation}
a_n=\frac{\alpha_n}{\sqrt{n}} \ \ \ \ \ \ a_n^{\dagger}=\frac{\alpha_{-n}}{\sqrt{n}}
\end{equation}
for $n>0$, this is different from quantum field theory in which creation and annihilation operators create and destroy particles. Here the operators create modes on the same existing string.\\
The vacuum state is defined to be annihilated by all annihilation operators (no modes of any kind exists)
\begin{equation}
\alpha^{\mu}_n|0>=\bar{\alpha}^{\mu}_n|0>=0 \ \ \ \ \ \ \ \forall \ \ n>0
\end{equation}
demanding normalized states we get 
\begin{equation}
<m|n>=\delta_{n,m}
\end{equation}
the general spate in the Fock space is then 
\begin{equation}
(\alpha_{-1}^{\mu_1})^{n_{\mu_1}}(\alpha_{-2}^{\mu_2})^{n_{\mu_2}}(\alpha_{-3}^{\mu_3})^{n_{\mu_3}}... (\bar{\alpha}_{-1}^{\nu_1})^{n_{\nu_1}}(\bar{\alpha}_{-2}^{\nu_2})^{n_{\nu_2}}(\bar{\alpha}_{-3}^{\nu_3})^{n_{\nu_3}}|0>
\end{equation}
where $n_{\mu_1}$ , $n_{\mu_2}$ ... are positive integers, this state corresponds to $n_{\mu_1}$ first modes and $n_{\mu_2}$ second modes and so on.\\
Now consider the state $|\psi>=\frac{1}{\sqrt{n}}\alpha^0_m|0>$ the inner product
\begin{equation}
    <\psi | \psi>=<0|\alpha^0_1 \alpha_{-1}^0|0>=<0|[\alpha^0_1 ,\alpha_{-1}^0]-\alpha^0_{-1} \alpha_{1}^0|0>
\end{equation}

the second term vanishes by the definition of the ground state, and by the commutation relations derived above
\begin{equation}
 <\psi | \psi>=-<0|0>=-1
\end{equation}
these states have negative norm, this is mathematically inconsistent so it is declared that these states correspond to unphysical states which must be removed form the theory, these states are called Ghosts.\\
To remove ghosts we need to modify the constraint (generators of Witt algebra) to its central extension called Virasoro algebra.\\ 
\subsubsection{\textbf{Witt algebra and Virasoro algebra.}}
we will prove that the set {$L_n$} forms an algebra called Witt algebra.\\
Define a multiplication operation to be the commutator, classically  these are numbers and functions so commute in the sense of quantum mechanical commutators but this is a general operation which in the classical case can be taken to be Poisson brackets. \\
Firstly we compute
\begin{equation}
[L_m,\alpha_n]=\frac{1}{2}\sum_p[\alpha_{m-p} \alpha_p , \alpha_n]
\end{equation}
after using the imposed conditions on the commutators (here we require the general operation to obey the same rules)
\begin{equation}
[L_n,\alpha_n]=-n\alpha_{m+n}
\end{equation}
then by the same procedure we get
\begin{equation}
[L_n,L_m]=(m-n)L_{m+n}
\end{equation}
so the set ${L_n}$ is closed under this operation, and it is easy to prove that it satisfies Bianchi identity, so it is a lie algebra called Witt algebra.\\
In the quantum theory this is correct except for the case $n+m=0$ we get normal ordering ambiguities (In quantum theory we must normal order the operators), its origin is a summation used to derive Witt algebra which is divergent in the quantum case. In this case we get an extra constant in $L_0$ 
\begin{equation}
L_0=\frac{1}{2}\sum_{- \infty}^{\infty}:\alpha_{-n} \alpha_n:=\frac{1}{2}(\alpha_0^2+\sum_{-\infty}^{-1}\alpha_{-n} \alpha_n+ \sum_{1}^{\infty}\alpha_{-n} \alpha_n)
\end{equation}
changing the sum index in the second term of eq.(72) by $n \rightarrow -n$
,we get
\begin{equation}
L_0=\frac{1}{2}\alpha_0^2+\sum_{1}^{\infty} \alpha_{-n} \alpha_{n},
\end{equation}
so we pick up a constant $\frac{1}{2} \alpha_0^2$ on normal ordering, this ambiguity tells us that we can not know which ordering is correct. To eliminate this ambiguity the constraint in the case of $L_0$ is modified to be $L_0-a=0$, where a is a constant to lift the ghosts, and the algebra is replaced by its unique central extension called Virasoro algebra (the rigorous definition of central extensions is found in Ref[12]) defined by
\begin{equation}
[L_n,L_m]=(m-n)L_{m+n}=\frac{c}{12}m(m^2-1)\delta_{m+n , 0}
\end{equation}
where c is the central charge defined by the operator power expansion of the stress tensor of the theory.\\
This is an example of an anomaly (a classical symmetry does not survive quantization). To remove the anomaly, c must equal zero.\\
Calculating c in bosonic string theory we get $c=-26$ and $a=1$, but knowing that c has two contributions one from the fields and the other from the ghosts, we need to add 26 scalar field to cancel the central charge (a scalar field has central charge =1), but each scalar field is associated with a degree of freedom i.e. a space time dimension, so the theory makes sense quantum mechanically only in 26 dimensions. Adding supersymmetry will add more ghosts of central charge 11, so the total central charge to be eliminated equal to -15
but here we add equal number of fermions($c=1/2$) 
$$D(1+1/2)=15 \rightarrow D=10$$
i.e. superstring theory makes sense in 10 space time dimensions.\\
\subsubsection{\textbf{D-branes}}
This is the main section in this part in which we use all previous sections to review some facts about D- branes.
D-branes are extended objects required by string theory (not put by hand) and it plays a central role in string cosmology.\\
D-branes are a consequence of imposing Dirichlet boundary conditions on open strings' endpoints, they are restricted to move in a lower dimensional "wall" which is a dynamical object evolving in time as well so it swipes a worldvolume of higher dimension. To construct the D-brane action in p dimensions (called Dp-brane) it is natural to begin with Nambu-Goto part describing the brane's motion
\begin{equation}
S=-T_p \int d^{p+1}\xi \sqrt{-det(\frac{\partial X^{\mu}}{\partial \xi^a}\frac{\partial X^{\nu}}{\partial \xi^b}g_{\mu \nu})}
\end{equation}
where $T_p$ is the tension of the brane and $\xi^a$ are the parameters parameterizing it and $g_{\mu \nu}$ is the space time metric.\\
Observing that $g_{\mu \nu}$ is the symmetric part dynamical field induced by $\alpha^i_{-1} \alpha^j_{-1}|0>$, the anti symmetric part $B_{\mu \nu}$ must contribute in the action as well making the action
\begin{equation}
S=-T_p \int d^{p+1}\xi \sqrt{-det(\frac{\partial X^{\mu}}{\partial \xi^a}\frac{\partial X^{\nu}}{\partial \xi^b}[g_{\mu \nu+\alpha' B_{\mu \nu}}])}
\end{equation}
the first approximation of this action gives a coupling with closed strings (gravity), when  a close string collide and interact with a brane it induces open strings to vibrate.\\
We want to do electrodynamics on D-branes, the suitable theory is Born-Infeld electrodynamics as discussed above. D-branes can have charge so there exist anti-branes and string charge explained in details in Ref[11], adding a term describing electrodynamics to the action we get
\begin{equation}
S=-T_p \int d^{p+1}\xi \sqrt{-det(\frac{\partial X^{\mu}}{\partial \xi^a}\frac{\partial X^{\nu}}{\partial \xi^b}[g_{\mu \nu}+\alpha' B_{\mu \nu}+\alpha'F_{\mu \nu}])}
\end{equation}
another part must be considered in the contribution of the dilaton field (the trace part of the state$\alpha^i_{-1} \alpha^j_{-1}|0>$) since it is a scalar it gives a coupling through it's vacuum expectation value(VEV) $g=e^<\phi>$ where $<\phi>$ is the VEV of the dilaton $\phi$ so we must add a coupling of $e^{\phi}$ in the action 
\begin{equation}
S=-T_p \int d^{p+1} \xi e^{\phi}  \sqrt{-det(\frac{\partial X^{\mu}}{\partial \xi^a}\frac{\partial X^{\nu}}{\partial \xi^b}[g_{\mu \nu}+\alpha' B_{\mu \nu}+\alpha'F_{\mu \nu}])}
\end{equation}
The final part is the coupling with supersymmetric fields or Wess-Zumino term, to the first approximation it can be written like any electromagnetism as a current multiplied by a gauge field, taking the field to be the supersymmetric tensor field $A_{\mu_1...\mu_p}$
and the current being produced by a p+1 dimensional brane we have
$j^{01...p}=\mu_p \delta^{(D-p-1)}(x-x(\xi))$ ,where $\mu_p$ is the charge density, leading to a term $\mu_p \int d^{p+1}\xi A_{01...p}(x(\xi))$, so the final action of a D-brane is
\begin{equation}
S=-T_p \int d^{p+1} \xi e^{\phi}  \sqrt{-det(\frac{\partial X^{\mu}}{\partial \xi^a}\frac{\partial X^{\nu}}{\partial \xi^b}[g_{\mu \nu}+\alpha' B_{\mu \nu}+\alpha'F_{\mu \nu}])}
\\ +\mu_p A_{01...p}(x(\xi))
\end{equation}
D-branes in string cosmology are where the universes are printed, and as discussed the coupling with closed strings and other branes influence cosmologies on these branes, cosmologies of this type are called brane world cosmologies to be discussed in part 3.\\
\subsubsection{\textbf{M-theory.}}
Incorporating supersymmetry in string theory gives rise to five different string theories, a detailed description can be found in Ref[11]. Fortunately the five theories are related by dualities namely S-duality and T-duality, forming a duality map between the theories, this gives rise to an idea that the five theories are just limits of one theory which was called M-theory, then it turns out that M-theory is just another limit of a yet unknown theory.\\
M theory is constructed as a strong coupling limit of type IIA string theory, an interesting feature of this limit is that one compact dimension unfolds giving rise to an 11 dimensional theory, it is used in string cosmology in some brane world cosmologies to host the vacuum states which represent the initial state of our universe. One useful state is called the Bogomol'nyi-Prasad-Sommerfield(BPS) state.\\
To define BPS states properly we need to know that type IIA string theory possesses a lot of supersymmetry (with supercharge=32), the D-branes solutions in this theory are invariant under half of these  supersymmetry generators, this symmetry appears also on the world volumes of the D-branes, this is an example of a BPS state. In general it is the state which preserves some supersymmetry i.e. invariant under some supersymmetry generators.\\
\subsubsection{\textbf{Horava Witten theory (Heterotic M-theory)}}
It was shown by Horava and Witten$[63,64]$ that the strong coupling limit to the $E_8 \times E_8$ heterotic string theory is M-theory in 11D compactified on $R^{10}\times S=R^{10} \times I$, so there is a duality between 10D vector multiplets on the boundaries of the manifold and the 11D supergravity multiplets in its bulk(ADS/CFT duality) with actions
\begin{equation}
S_{gauge}=-\frac{1}{4\lambda^2}\int_{M^{10}}d^{10} x \sqrt{g}tr[F^2]++...
\end{equation}
\begin{equation}
S_{supergravity}=-\frac{1}{2 \kappa^2}\int_{M^{11}}d^{11} x \sqrt{g} R+...
\end{equation}
where $\lambda$ is the gauge coupling constant, $\kappa$ is the gravitational coupling constant, R is the Ricci scalar, F is the field strength and the rest are irrelevant terms.\\
Another compactification on a 3D Calabi-Yau manifold gives a 4D supersymmetric theory with N=1 in the low energy limit. By energy comparison argument it was shown that this corresponds to a (4+1)D world bounded by two (3+1)D branes called the end of the world branes, these branes are fixed because it is the fixed points of the orbifold $S^1/Z_2$, by choosing the appropriate G-flux(for anomaly cancellation) and potential, the theory supports BPS branes, this is called Heterotic M theory or Horava-Witten theory.\\
The theory can contain the standard model's gauge group as the result of the breaking of one of the branes' gauge groups, this brane is called the visible brane (where our universe is located), the procedures of obtaining the standard model's gauge group is as follows:\\
In the compactification the standard embedding of the spin connection in the gauge connection is used leading to the group $E_8$ on one brane called the hidden brane and $E_6$ on the visible brane which upon breaking gives the standard model's gauge group(using the right G-instanton).\\
We can conclude that Horava-Witten theory can be a realistic model of particle physics, consequently it is used as the base for the ekpyrotic scenario to be discussed in part 3.\\
\subsubsection{\textbf{String thermodynamics and Hagedorn temperature.}}
Here we study the thermodynamics of a system of strings and define the Hagedorn temperature,the maximum temperature can be reached by these kind of systems(otherwise the partition function diverges).\\
Consider a system of non relativistic quantum strings with fixed endpoints, the Hamiltonian of the l-th mode is 
\begin{equation}
H_l=l \hbar \omega a^{\dagger}_la_l
\end{equation}
where $a_l$ is the annihilation operator of the mode l and $a^{\dagger}_l$ is the creation operator of the same mode.\\
The total Hamiltonian is the sum of all modes
\begin{equation}
H=\sum_{l=1}^{\infty}H_l=\hbar \omega \sum_{l=1}^{\infty}la^{\dagger}_la_l=\hbar \omega \hat{N}
\end{equation}
where $\hat{N}=\sum_{l=1}^{\infty}la^{\dagger}_la_l$ is the number operator.\\
From the Hamiltonian we can see that the energy of the system is $E=\hbar \omega N$ where N is an eigenvalue to the number operator, let $n_1,n_2,...$ be the number of modes in N i.e. how many first mode's creation operators in the number operator and how many second mode's creation operators and so on, and $P(N)$ be the number of ways to construct N (number of microstates as in the usual statistical mechanics), then the entropy is
\begin{equation}
S(N)=kln[P(N)]
\end{equation}
to get the value for ln[P] we calculate the entropy by another way and compare them, firstly we calculate the partition function
\begin{equation}
Z=\sum_{\alpha} e^{-E_{\alpha}\ kT}=
\sum_{ {n_i} }e^{-\hbar \omega / kT(n_1+2n_2+...)}=\Pi_{l=1}^{\infty}\sum_{n=0}^{\infty}e^{-\hbar \omega / kT (ln_l)}
\end{equation}
which is a geometric series so the result is
\begin{equation}
Z=\Pi_{l=1}^{\infty} \frac{1}{1-e^{-\hbar \omega l / kT}}
\end{equation}
so 
$$lnZ=-\sum_{l=1}^{\infty}ln[1-e^{-\hbar \omega l / kT}]$$
The free energy is then
\begin{equation}
F=-kTln[z]=kT\sum_{l=1}^{\infty}ln[1-e^{-\hbar \omega l / kT}]
\end{equation}
in the high energy limit i.e. $\hbar \omega / kT<<1$ converting the sum into an integral we get
\begin{equation}
F=kT\int_1^{\infty} ln[1-e^{-\hbar \omega l / kT}]dl
\end{equation}
computing the integral by substitution and expanding the ln we get
\begin{equation}
F=-\frac{\pi^2}{6 \hbar \omega \beta^2}
\end{equation}
where $\beta = \frac{1}{kT}$
then using $S=-\frac{\partial F}{\partial T}$ we get
\begin{equation}
S=\frac{k^2 \pi^2 T}{3 \hbar \omega}
\end{equation}
The energy is 
\begin{equation}
E=-\frac{\partial}{\partial \beta}ln[Z]=\frac{\pi^2}{6 \hbar \omega \beta^2}
\end{equation}
so
\begin{equation}
\frac{kT}{\hbar \omega}=\sqrt{\frac{6E}{\hbar \omega \pi^2}}
\end{equation}
substituting in the result for the entropy we get
\begin{equation}
S=\pi k \sqrt{\frac{2E}{3 \hbar \omega}}
\end{equation}
but $\frac{E}{\hbar \omega}=N$ so 
\begin{equation}
S=2\pi k \sqrt{\frac{N}{6}}.
\end{equation}
If we have b polarization directions then each part of N is partitioned into b parts leaving the total number bN so 
\begin{equation}
S=2\pi k \sqrt{\frac{bN}{6}}
\end{equation}
Comparing with $S=kln[P]$
we get $ln[P]=2\pi \sqrt{\frac{bN}{6}}$\\
In fact the real relation is 
\begin{equation}
p_b(N)=\frac{1}{\sqrt{2}}N^{\frac{-b-3}{4}}(\frac{b}{24})^{\frac{b+1}{4}}e^{2\pi \sqrt{\frac{bN}{6}}}
\end{equation}
and we derived the first approximation using classical arguments.\\
For bosonic strings $b=24$ because it is defined on 26D so the transverse directions are 24 so 
\begin{equation}
P_{24}(N)=\frac{1}{\sqrt{2}}N^{-27/4}e^{4\pi \sqrt{N}}
\end{equation}
This is the first ingredient in the analysis, the second one is to define the Hagedorn temperature, since the mass spectrum of a string is $M^2=\frac{1}{\alpha'}(N-1)\approx \frac{N}{\alpha'}$ for large N, where N is the number of transverse modes(in all the analysis the number of transverse mode is used), in the relativistic limit M=E so $\sqrt{N}=E\sqrt{\alpha'}$.\\
For bosonic strings $S=2 \pi k \sqrt{\frac{24N}{6}}$ as we derived, substituting about N we get
$$
S=4 \pi k E \sqrt{\alpha'}
$$
but 
$$
\frac{1}{T}=\frac{\partial S}{\partial E}=4 \pi k  \sqrt{\alpha'}$$

this gives a special value for the temperature called the Hagedorn temperature
\begin{equation}
T_{H}=\frac{1}{4 \pi k \sqrt{\alpha'}}
\end{equation}
define the factor 
\begin{equation}
\beta_H=\frac{1}{kT_H}=4 \pi \sqrt{\alpha'}.
\end{equation}
The third step is recalling that the partition function for a relativistic particle in a box is 
\begin{equation}
Z_p=Ve^{-\beta m}(\frac{m}{2 \pi \beta})^{d/2}
\end{equation}
where V is the volume of the box, d is the number of spacial dimensions and m is the mass of the particle.\\
Using the definition of the Hagedorn temperature we get $4 \pi k T_{H} \sqrt{\alpha'}=1$
so
$$\frac{m}{2 \pi \beta}=\frac{m}{2 \pi \beta}4 \pi k T_{H} \sqrt{\alpha'}=2 \sqrt{\alpha'}m KTKT_H$$
$$\beta m= \beta m 4 \pi k T_{H} \sqrt{\alpha'}=4 \pi \sqrt{\alpha'}m \frac{T_H}{T}$$
substituting in the partition function with $d=25$ (bosonic string theory case), we get
\begin{equation}
Z_p=V 2^{25/2}(\sqrt{\alpha'}m KTKT_H)^{25/2}e^{-4 \pi \sqrt{\alpha'} m \frac{T_H}{H}}.
\end{equation}
The final step is to apply all this on a single open string.\\
The states of such a string is given by
\begin{equation}
|\lambda,P>=\Pi_{n=1}^{\infty}\Pi_{I=2}^{25}(a_n^{I\dagger})^{\lambda_{n,I}}|p^+,p_T>
\end{equation}
where $\lambda_{n,I}$ are integers define how many creation operators are there for a given n and I(n is the mode and I labels the transverse directions),$p^+$ is the momentum in the light cone coordinates defined by$p^+=\frac{p^0+p^1}{\sqrt{2}}$ and$p_T$ is the momentum in the same coordinates but the transverse directions.\\
Since each mode corresponds to a relativistic particle then the total partition function is nothing but the sum of all partition functions of the modes
$$Z=\sum_{\lambda_{n,I}Z_p(m^2)}$$
and since the mass depends only on the number of the mode we can change the sum to a sum on N
$$Z=\sum_{N=0}^{\infty}Z_p(N)$$
but for each N there are many "microstates" of the system as discussed before, so the number of such states must be multiplied by the state's partition function to give the total partition function on adding 
\begin{equation}
Z=\sum_{N=0}^{\infty}P_{24}(N)Z_p(N)
\end{equation}
for the approximations in the formula for $P_{24}$ to hold we must have large N, so separate the sum into two parts the first part contains small N values which we can not approximate to an integral or use our formula for $P_{24}$, call it $Z_0$, and the other part beginning with the $N_0$ after which the approximations hold and then we can approximate the sum into an integral giving
$$Z=Z_0+\int_{N_0}^{\infty}P_{24}Z_p(N)$$ 
substituting by the formulae of $P_{24}$ and $Z_p$ we get
\begin{equation}
Z=Z_0+2^{13}V(KTKT_H)^{25/2}\int_{M_0}^{\infty} d(\sqrt{\alpha'}m)e^{-4 \pi \sqrt{\alpha'}m(\frac{T_H}{T}-1)}
\end{equation}
which diverges for any temperature above the fixed Hagedorn temperature, this says that in strings systems the temperature has an upper bound like the velocity of light being the upper bound of the velocities in the theory of relativity. This fact is crucial in some string cosmology models like string gas cosmology in part 3.\\
\section{\textbf{Part 3 : String Cosmology}}
In the main part of the review models of string cosmology is discussed model by model. String cosmology originated because to test string theory a very high energy is needed and we have no technology to reach it at the moment or even in the near future, but events like supernovae explosion and some other cosmic events can in fact be of very high energy so can serve as a possible method of testing string theory, another reason is that string theory requires gravity and GR can be derived form string theory as a first approximation, this means that string theory must have the ability to explain cosmology and deal with the universe.\\
Some string cosmology models tried to predict inflation like brane-antibrane inflation but there were some problems in string inflation as discussed in Ref[62] that brane inflation in flat spacetime is impossible due to the steep potential induced between the branes, however, some models overcame this problems as we will see, other introduce an alternative to inflation like the ekpyrotic theory, the two categories are discussed in this part.\\

\subsection{\textbf{Brane Inflation}}
The simplest brane world cosmological model is the brane inflation$[63]$, in this model and all brane world models we have a higher dimensional space where the lower dimensional D-branes are embedded. The higher dimensional space between D-branes is called the bulk, so as discussed in part 2 open strings are confined and restricted to the D-branes, but the closed strings can move through the bulk. In string theory the standard model particles are open string modes so are confined to a D-brane i.e. our universe is printed on a D-brane of three non compact spacial dimensions. 
In this model a regime is proposed to induce inflation through a system of D-branes .\\
Suppose the simplest system of two D-branes in the BPS state, the force between them can be generated by the exchange of closed string states(gravitons) i.e. by gravitational force, by the exchange of massive bulk modes, by the tension of open strings extends from one brane to the other and finally by the RR antisymmetric charge (it is assumed that that the branes have the same charge). If the two branes coincide another force arise from the interaction between open string modes on the branes, in the case of BPS branes the total force or the total vacuum energy vanishes$[66]$, in the case of displaced branes system after supersymmetry breaking the cancellation of forces does not hold anymore and an attractive force is induced.\\
The remaining ingredient is the inflaton field, if it was chosen to be a field living on a D-brane, some serious problems arise as in Ref[67] this motivates the choice of the inflaton to be a scalar field whose vev is the separation between the two D-branes i.e. the separation is the field itself (which is a modulus of course), this inter-brane mode is weakly coupled to bulk modes, and induces a slow roll inflation as we will see, when the branes are close to each other the inflaton couples with the modes on the other brane inducing a reheating on the brane but not in the bulk due to the weak coupling with the bulk modes, after the collision the branes oscillate around the equilibrium point giving rise to our universe on one of the branes.\\
The proposed potential at large distances is
\begin{equation}
V(r)=T(\alpha-f(r/r_0)+b_i \frac{e^{-m_ir}}{r^{N-2}}+\frac{c}{r^{N-2}}+kr)
\end{equation}
with $\phi \approx M^2_{pl}r$
where $\phi$ is the inflaton field, $M_{pl}$ is the planck mass, T is the brane tension (the two branes are assumed to have the same tension), r is the reparation between the two branes, $\alpha$ , $b_i$, c are model dependent constants ,$r_0$ is the brane thickness (beyond which the localized open string modes deacy exponentially), k is proportional to the stretched strings density, N is the extra dimensions of the bulk and $m_i$ are the masses of the bulk modes. The first two terms come from the modes localized on each brane, the third term originates from the massive bulk modes, that is why there is a decaying exponential is a term in the Yukawa potential form, the fourth term is the massless particles exchange potential and the last term came from the strings stretched from one brane to the other (a potential of an elastic string is proportional to the distance and the number of strings per unit volume which is exactly the last term).\\
Writing the potential in terms of the inflaton and the mass parameter $M_i=M^2_{pl}/m_i$ and $M \approx T^{1/4}$ we get
\begin{equation}
V(\phi)=T(\alpha-f(\phi/M)+\frac{1}{|\phi|^{N-2}}[\beta_ie^{-|\phi|/M_{i}}+\gamma])
\end{equation}
where $\beta_i$ and $\gamma$ are constants.\\
taking $\gamma=0$ and the Hubble parameter $H^2=\frac{T\dot{r}+V}{3M_{pl}^2}$ and the equation of motion (Friedmann equations) for an attractive potential
\begin{equation}
\ddot{\phi}+3H\dot{\phi}+V'=0
\end{equation}
where the dot is the derivative with respect to time and the prime is the derivative with respect to $\phi$.\\
The potential satisfies the slow roll conditions if we assume $V<<\dot{r}$, the end of the inflation is characterized by the breaking of one of the two conditions, which is at$\phi_{end}$ such that
\begin{equation}
\frac{\beta}{\alpha \phi_{end}^{N-2}}e^{-\frac{\phi_{end}}{M}} \approx (\frac{M}{M_{pl}})^2
\end{equation}
it is shown explicitly in Ref[63] that this gives the right number of e-folds required for the inflation we know.\\

\textbf{Summary}: The system begins with two D-branes far away form each other, on supersymmetry breaking an attractive force is generated attracting the two branes slowly towards each other inducing an inflation on one of the branes, when the branes come close the localized modes on them interact but still early interacting with the bulk modes and the potential becomes steep so inflation ends, then on collision and brane oscillations around the equilibrium point the energy of the collision reheats the branes mostly in the form of radiation (not the bulk due to weak interaction with bulk modes), then the radiation dominated era begins as in standard cosmology.\\

\textbf{Open Problems}\\

Although this model gives rise to the slow roll inflation which solves the flatness, horizon and unwanted relics problems.\\
1- the model gives no precise mechanism to end inflation, and no details on what happened after the collision.\\
2- the model does not provide solutions to new problems other than the standard cosmology problems.\\
3- the most serious problem is that this model is far from rigorous and is based only on physical intuition, the potential is not derived from string theory but rather guessed making a huge computability problem (we can not progress further because many things are assumed or guessed not derived from an underlying theory).\\
D-brane inflation is reviewed in Ref[80] and Ref[71].\\

\subsection{\textbf{Brane-Antibrane Inflation}}
The brane-antibrane inflation model is a similar model to the brane-brane inflation but the system begins with a brane and an antibrane (a brane with opposite RR charge), this induces a natural attractive force due to the exchange of massless bulk modes, the first model of this type assumes a brane-antibrane system where the brane and the antibrane are parallel to each other and are far apart, their interaction is well understood from string theory$[66]$, the brane and the antibrane are assumed to approach each other slowly causing a slow roll inflation like the brane-brane model, however this attempt failed because the interaction derived from string theory is strong so the slow roll can not occur as we will prove.\\
Consider the system described above, the total effective action is the sum of three parts the bulk action
\begin{equation}
S_B=-\int d^4xd^dy \sqrt{-g}[\frac{M_s^{2+d}}{2}e^{-2\phi}R+...]
\end{equation}
where x are the four spacetime dimensions, y are the extra dimensions' coordinates, g is the determinant of the metric tensor, $M_s$ is the fundamental string scale , R is the Ricci scalar, $\phi$ is the dilaton field and the rest are irrelevant terms due to the bulk modes.\\ 
The brane and the antibrane action by expanding Born-Infeld action 
\begin{equation}
S_{b_i}=- \int d^4x d^{p-3}y \sqrt{-\gamma}[T_p+...]
\end{equation}
there $i=1,2$ for $b_1$ is the brane and $b_2$ is the antibrane, $\gamma$ is the induced metric discussed in part 2 and $T_p=\alpha M_s^{p+1}e^{-\phi}$ is the brane tension($\alpha$ is a dimensionless constant).\\
Adding the brane and the antibranes actions in the center of mass coordinates x,y we get
\begin{equation}
S_{b_1}+S_{b_2}=- \int d^4x d^{p-3}y \sqrt{-\gamma} T_p[2+1/4g_{mn}\gamma^{ab}\partial_a y^m \partial_b y^n+...]
\end{equation}
The authors of Ref[66] made some assumptions for their analysis to hold:\\
1- The moduli are stabilized by some unknown mechanism, so we can use the volume moduli $V_{\perp}=r_{\perp}^{d-p+3}$(the volume transverse to the brane), and $V_{//}=r_{//}^{p-3}$ (the volume along the brane) as parameters and they chose them to satisfy 
\begin{equation}
M_p^2=e^{-2\phi}M_s^{2+d}V_{\perp}V_{//}
\end{equation}
where $M_p\approx 10^{18}$.\\
2- The volumes in the units of the fundamental string scale is very large $M_s r_{\perp}>>1$ and $M_s r_{//}>>1$, this is required to treat bulk dynamics with low energy effective field theory in string theory.\\
With these assumptions the potential as derived in Ref[15] is
\begin{equation}
V(y)=\frac{2 \alpha e^{\phi}}{(M_s r_{\perp})^{d_{\perp}}}M_s^2 M_p^2- \frac{1}{y^{d_{\perp}-2}}\frac{\alpha^2 \beta e^{\phi}M_p^2}{M_s^{2(d_{\perp}-2)}r_{\perp}^{d_{\perp}}}
\end{equation}
this is valid for $M_s^{-1}<<y<<r_{\perp}$.\\
Checking the slow roll conditions using the same procedure as the brane-brane inflation model, the second condition reads
\begin{equation}
\eta \approx -\beta (d_{\perp}-1)(d_{\perp}-2)(\frac{r_{\perp}}{y})^{d_{\perp}}
\end{equation}
but the assumption is $r_{\perp}<<y$ so $|\eta|<<1$ do not hold, so the model fails to give a slow roll inflation.\\
The natural question to ask is what if we relaxed the condition $y<<r_{\perp}$? does this induce a slow roll inflation?\\
the answer is yes, consider a brane-antibrane system on the torus $(R/Z)^{d_{perp}}$ which is equivalent to a lattice as in the figure 5 below
\begin{figure}[hbtp]
\centering
\includegraphics[scale=0.25]{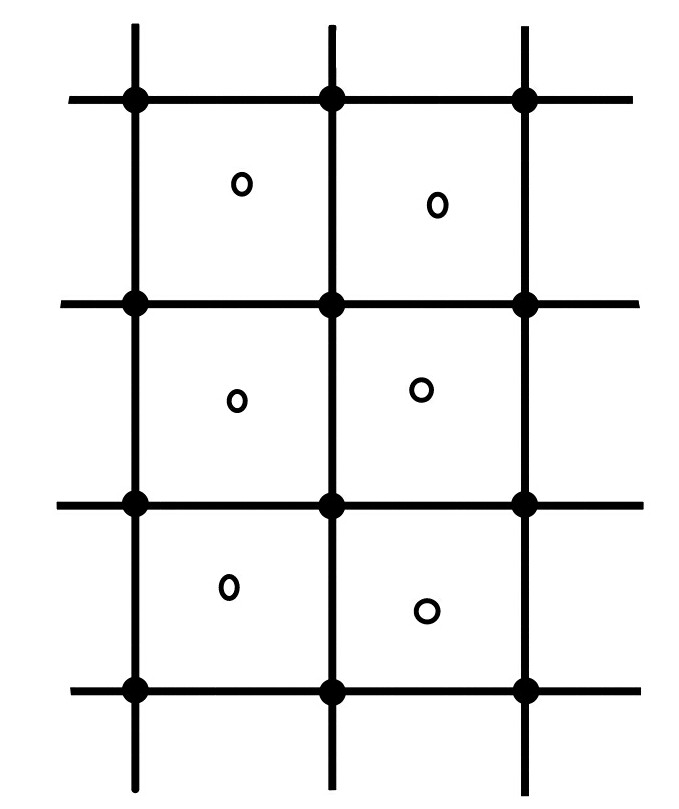}
\caption{The lattice equivalent to $(R/Z)^{d_{\perp}}$ of the brane-antibrane system where the solid circles represent the brane(and its images) and the open circles represent the antibrane(and its images) near the center of the cell$[71]$.}
\end{figure}
and the potential is 
\begin{equation}
V(r)=\frac{2 \alpha e^{\phi}}{(M_s r_{\perp})^{d_{\perp}}}M_s^2 M_p^2-\sum_i \frac{1}{|r-r_i|^{d_{\perp}-2}}\frac{\alpha^2 \beta e^{\phi}M_p^2}{M_s^{2(d_{\perp}-2)}r_{\perp}^{d_{\perp}}},
\end{equation}
where $r_i$ is the position of the branes with respect to some origin and $r$ is the position of the antibrane (it is sufficient to consider one because the images do not contribute to the potential).\\
By symmetry arguments the first and the second derivatives of the potential vanish, studying the antibrane motion due to a displacement z from the center of the hyper-cubic cell, expanding the potential in terms of a power series of z, the expansion can not have odd powers due to reflection symmetry, nor a quadratic term due to the vanishing second derivative, so the potential can be approximated as
\begin{equation}
V(z)=\frac{2 \alpha e^{\phi}}{(M_s r_{\perp})^{d_{\perp}}}M_s^2 M_p^2-1/4 \gamma M_s^{d+2} e^{2\phi} T^2_p V_{//} r_{\perp}^{-(d_{\perp}+2)} z^4
\end{equation}
the second slow roll condition is 
\begin{equation}
\eta \approx -3 \gamma (\frac{z}{r_{\perp}})^2<<1
\end{equation}
which is true because z is very small compared to $r_{\perp}$.\\
The most interesting part is that string theory provides a mechanism to end inflation$[16]$ in this case as follows: the potential is attractive and is valid if the separation is much larger than the string scale, when the brane and the antibrane approach each other so that the separation is comparable with the string scale the approximation breaks down, the conjectured exact tachyonic action is 
\begin{equation}
S_T=-M_s^2 M_p^2 \int d^4x e^{-|T|^2}[1+\kappa_1(\frac{2+\kappa_2 |T|^2}{M_s^2})|\partial T|^2]
\end{equation}
where $\kappa_1$ and $\kappa_2$ are dimensionless constants, the detailed behavior of the action is discussed in Ref[15], in this case the massive string modes become massless then tachyonic which ends the inflation.\\
Afterwards by sen's conjecture$[68]$ the height of the tachyon potential equals the brane tension, and the result of the collision is closed string modes, but if tachyons have non trivial winding stable lower dimensional branes and antibranes can be produced.\\
This model implies a cascade to lower dimensional brane-antibrane systems, in the case of $p+1$ dimensional world we begin with Dp-$\bar{Dp}$ branes system, on colliding they give unstable non-BPS $D(p_1)$ branes which in turn decay to $D(p-1)-\bar{D(p-2)}$ system and repeat(for each collision reheating occurs due to the release of heat in form of radiation mainly as the result of the collision and the decay process). This is true for all brane configurations in type I and type IIB superstring theories. This cascade stops when branes and antibranes can not find each others to collide, the estimate$[69,70]$ is that in a d spacial dimensional space it is unlikely for branes of dimension less than or equal p/2 to collide, so in the case of our 10 dimensional world or 9 spatial dimensional world the cascade stops when the branes have three spatial dimensions.\\

\textbf{Summary}: The old brane antibrane idea in which one brane and one antibrane interact can not give slow roll inflation, so a new idea arise to discuss the same system but on a square torus which indeed gives a slow roll inflation, according to this model we begin with a 9 dimensional space filled with a gas of 9 dimensional branes and antibranes, these collide to give a gas of 7 dimensional branes and antibranes with a reheating and so on until the system is 3 dimensional so branes and antibranes are unlikely to meet and collide so our universe is 3 dimensional.\\

\textbf{Open Problems}:\\

1- it is assumed that the 4 dimensional effective FRW background is valid which is not true in general.\\
2- the most serious problem is that the assumption that the moduli are stabilized by some unknown mechanism, this is not an easy thing to assume making the model derivation nearly assumed not derived from an underlying theory which is the same problem as the brane-brane inflation but less serious.\\
3- the model does not provide a detailed mechanism for reheating.\\
4- the model has no realistic D brane models in which the standard model particles live (intersecting D branes discussed in Ref[61]) and Ref[71].\\
This model is reviewed in Ref[62] and Ref[71].\\

\subsection{\textbf{The ekpyrotic and cyclic scenarios}}
So far the models reviewed aims to recreate slow roll inflation by using objects and frameworks from string theory, the ekpyrotic and the cyclic model unlike the others give an alternative to inflation allowing the description of our cosmology without any super laminar expansion of the universe.\\
The ekpyrotic scenario$[72]$ uses the Horava Witter theory discussed in part 2 to set up a cosmological model, the model resembles the previous ones in that the universe is printed on a D-brane embedded in a higher dimensional world, and that the separation between branes is identified with the inflaton field but here with a reason not just an assumption and that the big bang resulted from a collision between the branes, However the model proposed solutions to the flatness,horizon and unwanted relics problems without inflation as follows.\\
The framework used is M-theory compactification $S^1/Z_2$ in which the endpoints have gauge group $E_8$, compactifying the result space on 6 dimensional Calabi Yau manifold gives two D4 branes in a 5 dimensional bulk, one brane is called the visible brane on which our universe is printed, the other is called the hidden brane, the two are called the end of the world branes. In principle in this regime more branes are present but they can move through the bulk so they are called bulk branes. It is assumed that the bulk brane is much lighter than the end of the world branes to consider any back reaction during the collision as a small perturbation.\\
The authors$[73,74]$ proposed that a bulk brane moved slowly from the hidden brane under a potential of the form $V=-V_0 e^{-c\phi}$ where $\phi$ is the inflaton field(the distance between the bulk brane and the visible brane) and $c>>1$,$V_0$ are constants in what called the ekpyrotic phase (contacting universe) and as the separation become close to plank's scale the potential becomes irrelevent and here string theory effects become important in what called the kinetic phase, then the bulk brane collides with the visible brane where the kinetic energy is converted to radiation on the brane just as the brane-brane inflation model but the difference is that this identification is due to the choice of the metric of the bulk to have a warp factor so the bulk brane moving from small curvature to high curvature meaning that the scale factor depends on the distance(it depends on the curvature which depends on the distance), then it is convincing to interpret this distance as the vev of the inflaton. To induce a flat universe the branes must be nearly BPS and parallel to each other, during the bulk brane movement it gains kinetic energy due to non perturpative effects mainly gravity which is essential to induce a hot big bang and begin a radiation dominated era. Another achievement done by this model is that the density perturbations arise naturally due to quantum fluctuations in the bulk brane causing regions to collide before another regions, the regions collided earlier are reheated earlier so cooled earlier inducing inhomogeneities required for large scale structure formation.\\
The flatness problem is solved by construction, the horizon problem is solved by observing that two events separated by more than several Hubble's radius are causally connected here due to the common causal link between them (the collision) i.e.the CMB looks alike on the large scale because it originated from the same collision with the same conditions. The unwanted relics problem is simple in this scenario because these relics require a certain high temperature to be created, if we assume that the collision temperature is lower than this known temperature the problem is completely avoided.\\
Mathematically the model has gravity and a scalar field so the Lagrangian density is 
\begin{equation}
\mathcal{L}=\sqrt{-g}[R-1/2\partial_{\mu}\phi \partial^{\mu}\phi -V(\phi)]
\end{equation}
where R is the Ricci scalar and g is the determinant of the metric tensor.\\
To describe the ekpyrotic phase we need the KG equation for the scalar field and the Friedmann equations which are
$$\ddot{\phi}+3H\dot{\phi}+V=0$$
$$3H^2=1/2 \phi^2+V$$
$$\dot{H}=-1/2\phi^2$$
with solution
$$a=(-t)^{2/c^2}$$
$$\phi=\frac{2}{c} ln[-tc \sqrt{\frac{V}{2}}]$$
comparing with the general formula for a in part 1 we get 
\begin{equation}
w=c^2/3-1>>1
\end{equation}
i.e. a contracting phase.
The kinetic phase is represented by neglecting the potential this gives us the solutions for the stiff matter as in the standard cosmology.\\
A serious problem arises is that this model violates the null energy condition to see this we compute $\dot{H}=-1/4(\rho+P)$
and we know that $P=1/2 \dot{\phi}^2-V$ and $\rho = 1/2 \dot{\phi}^2+V$
so $\dot{H}=-1/4\dot{\phi}^2<0$ i.e. H is monotonically decreasing.\\
The problem is that in the standard cosmology $\dot{H}>0$ which is impossible to achieve since H is negative and monotonically decreasing.\\
To solve this problem a modification is proposed to the ekpyrotic scenario in which the end of the world brane are themselves colliding, here we have a singularity corresponding to the moment when the size of the extra dimension is zero after which the branes pass through each other and the separation increases again increasing the scale factor which is our standard cosmology, here the null energy condition is satisfied(we have a singularity which can make the transition from negative to positive H).\\
This naturally motivates a third model,the cyclic model, in which after the collision of the end of the world branes and their separation increase, they attract each others again in a potential similar to a finite well and the cycle repeats itself forever (ekpyrotic phase then kinetic phase then a collision then a separation then another ekpyrotic phase and repeat).\\
Recently, This model was used to eliminate the singularity in black holes$[75]$. The construction proposes that a new emergent univesr emerges inside the black hole's horizon which was in an ekpyrotic phase and now in the expanding phase. The theory is based on introducing S-branes$[76]$ to the theory. S-branes have zero energy density, and negative pressure, Thus, it allows bounces without singularities.\\
This theory automatically solves the information paradox of black holes since information falling into the black hole is not lost but goes to the other universe$[75]$.\\
Cosmological perturbations and a mechanism of reheating were also studied for the emergent universe$[77,78]$. The radiation dominated era and the hot big bang were achieved by the decay of the S-brane after the bounce. Further developments in the theory including the interaction between the emergent universe and the Hidden brane is still in progress.\\

\textbf{Summary}: According to the ekpyrotic theory the world is five dimensional with one compact dimension bounded by two heavy end of the world branes and between them the light bulk branes which are allowed to move through the bulk, one of the bulk branes moved from the hidden to the visible brane and collide with it making a hot big bang followed by radiation dominated era, but this model failed due to the violation of the null energy condition so they got rid of the bulk brane to get a singularity.\\
The cyclic model is similar but with a periodic nature of attraction between the branes causing an endless cycle of contractions and expansions.\\
A comparison between the standard inflation theory and ekpyrotic/cyclic theories is presented in Ref[81]\\

\textbf{Open Problems}\\

1- To solve the flatness and the horizon problems they assumed the branes to be parallel which is a huge fine tuning.\\
2- Until now there is no quantitative calculations are done to compute the temperature of the collision to prove it is below the temperature required to create monopoles and other topological defects, it is just assumed to be less.\\
3- Like all the previous models the authors assume implicitly that the moduli of the CY manifold are stabilized.\\
4- Like the bran-brane inflation the potentials proposed to be working well are guessed not derived from a theory.\\

\subsection{\textbf{The new ekpyrotic model}}
The problem in the ekpyrotic scenario which lead to its modification is the violation of the null energy condition leading to severe instabilities such as ghosts of arbitrary large mass, this  was solved by the modification addressed earlier or the cyclic model. The new ekpyrotic model$[79]$ solved the problem with a different approach, in this model the authors propose a way to violate the null energy condition by ghost condensation with higher derivative kinetic terms so that the solution is ghost free, the idea is not to evade the violation of the null energy condition itself but to solve the instability problem caused by such violation.\\
The setup is the same as the ekpyrotic model but here a new phase is proposed called the ghost condensate phase followed by the collision, The higher derivative kinetic term used to describe a scalar field in the ghost condensate phase is
\begin{equation}
\mathcal{L}= \sqrt{-g} P(X)
\end{equation}
where P is an arbitrary function and
\begin{equation}
X=-\frac{1}{2m^4}(\partial \phi)^2
\end{equation}
the equation of motion is 
\begin{equation}
\frac{d}{dt}(a^3 \frac{dP}{dX} \dot{\phi})=0
\end{equation}
we require P to have a minimum (ghost condensation point) without loss of generality by rescaling and redefining the field we choose the minimum to be at $X=1/2$, the exact solution for the equation of motion is
\begin{equation}
\phi=-m^2 t 
\end{equation}
 the pressure and the energy density are
 \begin{equation}
 P=M^4 P(X) 
 \end{equation}
 \begin{equation}
 \rho = M^4(2\frac{dP}{dX}-P(X))
 \end{equation}
 where M is the scale determined by the underlying theory.\\
The aim is to compute the pressure and the energy density near the minimum, so a perturbation is added to the field giving 
\begin{equation}
\phi=-m^2 t +\pi(x,t)
\end{equation}
where the function $\pi$ is the perturbation, by taylor expansion to the second order
\begin{equation}
X=1/2-\frac{\dot{\pi}}{m^2}+\frac{\dot{\pi}^2-(\nabla \pi)^2}{2m^4}
\end{equation}
\begin{equation}
\frac{dP}{dX}=-\frac{dP}{dX^2}(1/2)\frac{\dot{\pi}}{m^2}.
\end{equation}
We note that this minimum has no ghosts as if we substitute in the Lagrangian the leading order in $\dot{\pi}$ has a negative sign i.e. a stable equilibrium with no ghosts or  unstabilities.\\
Adding a potential, the pressure and the energy density are given by
$$P=-V$$
$$\rho=-\frac{KM^4\dot{\pi}}{m^2}+V$$
where K is the second derivative of P with respect to X evaluated at the minimum.\\
From this we get
$$\dot{H}=\frac{KM^4\dot{\pi}}{2M_{pl}^2m^2}$$
this can be expansion or contraction depending on the sign of $\dot{\pi}$.\\
The forms of $P(X)$ can be guessed easily form the information we have (steep in the ekpyrotic phase and in the ghost condensate phase it has a minimum at $X=1/2$), the guess is
\begin{equation}
M^4P=m^4X
\end{equation}
in the ekpyrotic phase and 
\begin{equation}
P=\frac{K}{2}(X-1/2)^2
\end{equation}
in the ghost condensate phase.\\
The question here is which phase precedes the other as X increases, if the ekpyrotic phase came first there would be an increase in P then a minimum this implies that there is a maximum in between, this corresponds to a real (perturbative) ghosts which is bad for a model, so we are left with the second possibility that is the ekpyrotic phase came later.\\
The natural question to ask is how can we confine the field to be in the vicinity of the ghost condensate point(the minimum)?\\
the answer is by choosing a potential to do so, the potential proposed is 
\begin{equation}
V\approx \alpha \Lambda^4(1-\beta \frac{\Lambda^2 \phi}{m^2 M_{pl}}),
\end{equation}
with $\alpha$ and $\beta$ are positive (proved by requiring that we have the null energy condition satisfied in the beginning and then violated) and $\Lambda$ is some scale.\\
As any model there are some approximations and conditions must hold for the model to be consistent. The approximation required to hold throughout the null energy condition violating phase is $$\dot{\pi}<<m^2$$
this is obvious as $\pi$ is a small perturbation by definition, the implications on the kinetic and potential functions are discussed in details in the original paper defining the model.\\
The condition must hold come from Friedmann equations which imply that Hubble parameter is directly proportional to the square root of the potential, applying this to the minimum and the ekpyrotic potentials, and relating them to the number of the e-folds discussed in part 1 we get
$$e^N=\frac{H_{min}}{H_{ek}}=\sqrt{\frac{V_{min}}{V_{ek}}}$$
where the subscript "ek" to a function refers to the function at the ekpyrotic phase, this implies
$$|V_{ek}|=e^{2N}|V_{min}|$$
The lower bound of $|V_{min}|$ is derives as follows: we need P to be approximated as linear so $X=\dot{\phi}^2$ must be much larger than $m^4$ so that its contribution is negligible, but in the scaling phase $\dot{\phi}^2 \approx -2V$ so as a special case $|V_{ek}|>>m^4$ but using$|V_{ek}|=e^{2N}|V_{min}|$ we have our bound
\begin{equation}
m^4 e^{2N}<<|V_{min}|.
\end{equation}
The upper bound is derived from the scaling solution and after simple algebra we get
$$|V_{min}|<<\frac{M^4K}{p}$$
where $p=Ht<<1$,
so the condition on the potential minimum is
\begin{equation}
m^4 e^{2N}<<|V_{min}|<<\frac{M^4K}{p}
\end{equation}
There is another model proposed uses the same mechanism to induce inflation called the Ghost inflation model using the method presented in Ref[86] to generate ADS spaces\\
An ongoing current research aims to incorporate the construction in Ref[75] discussed in the previous section to this model. The relation between the emergent universe and the Hidden brane is yet unknown.\\

\textbf{Summary}: This model is the same as the ekpyrotic model but instead of searching for a way to satisfy the null energy condition by imposing a setup in which a singularity occurs, here a higher derivative term is added to evade the problem of stability itself even if the null energy condition is violated. The new phase proposed is the ghost condensate phase as a part of the kinetic phase in which the higher derivative terms function has a minimum. To keep the field in its vicinity a scalar potential is assumed to have a low minimum with steep nature so the field loses and gains kinetic energy in the way that it is kept near the minimum of the higher derivative function, this minimum has no ghosts so the instability problem is solved.\\ 

\textbf{Open Problems}:\\

Beside the problems of the ekpyrotic scenario, this model has an additional problem is the presence of gradient instabilities, this may be solved by adding more higher derivatives.\\
\subsection{\textbf{String gas cosmology}}
String gas cosmology(SGC) is a vast subject with many literature and many results, the most significant results and equations are reviewed in this section.\\
String gas cosmology first introduced by Brandenberger and vafa$[82]$ introduces an alternative to inflation, The setup proposed is that the universe begin with all the dimensions are compact and filled with a gas of fundamental strings, then three dimensions only are freed to expand by a mechanism discussed later.\\
The model is based on four assumptions$[82]$:\\
1- The background fields are all assumed to be homogeneous, however, a generalization is made to non homogeneous fields.\\
2- The higher derivative derivative terms ,like the ones discussed earlier in the other models, can be ignored, this is due to the assumption that the fields are evolving slowly with time (Adiabatic approximation).\\
3- The strings are weakly coupled as we are working with fundamental strings.\\
4- All spatial dimensions are assumed to be toroidal, this assumption is also relaxed in later literature.\\
The universe was assumed to begin in a string scale dimensions near the Hagedorn temperature called the Hagedorn phase with the T-duality is respected, the Polyakov action in this case (string in a time dependent background) is
\begin{equation}
S=-\frac{1}{4\pi \alpha'} \int d^2 \sigma [\sqrt{-\gamma} \gamma^{ab}g_{\mu \nu}\partial_a X^{\mu} \partial_b X^{\nu} + \epsilon^{ab} B_{\mu\nu} \partial_{a}X^{\mu} \partial_b X^{\nu} + \alpha' \sqrt{-\gamma}\phi R^{(2)}],
\end{equation}
where $g_{\mu \nu}$,$B_{\mu \nu}$ and $\phi$ are the graviton , the Kalb-Ramond field and the dilaton respectively(the symmetric,antisymmetric and the trace part of a generic second rank tensor representing fields in string theory),$\gamma^{ab}$ is the auxiliary metric ,$\gamma=det[\gamma^{ab}]$and R is the Ricci scalar of the world sheet of the strings.\\
The equations of motion for the fields X are
\begin{equation}
\partial_a(\sqrt{-\gamma}\gamma^{ab}\partial_bX^{\mu})+\Gamma^{\mu}_{\lambda \nu}\sqrt{-\gamma}\gamma^{ab}\partial_bX^{\lambda}\partial_bX^{\nu}+1/2H^{\mu}_{\lambda \nu} \epsilon^{ab} \partial_aX^{\lambda} \partial_b X^{\nu}=0
\end{equation}
with the constraints coming from varying the action with respect the auxiliary metric $\gamma^{ab}$
\begin{equation}
g_{\mu \nu}[\partial_a X^{\mu}\partial_b X^{\nu}-1/2 \gamma_{ab}\gamma^{cd}\partial_c X^{\mu}\partial_d X^{\nu}]=0
\end{equation}
where $H^{\mu}_{\lambda \nu}$ is the field strength and is given by the exterior derivative of the Kalb-Ramond field $H=dB$.\\
The cosmological ansatz is 
\begin{equation}
ds^2=-dt^2+\sum_{i=1}^d a_i(t)dx_i^2
\end{equation}
and $a(t)=e^{\lambda(t)}$,$\phi=\phi(t)$, $B=0$. 
Fixing the gauge of the auxiliary metric $\gamma_{ab}=f(\sigma,\tau)\eta_{ab}$ and using the adiabatic approximation, the equations of motion can be written as
\begin{equation}
(\partial_{\sigma}^2-\partial_{\tau}^2)X^{\mu}(\sigma,\tau)=0
\end{equation}
which has the same form of solution as the free strings discussed in part 2 except that
\begin{subequations}
\begin{align}
p_R^m=\frac{\sqrt{\alpha'}}{R}n^m-\frac{R}{\sqrt{\alpha'}}w^m \\
p^m_L=\frac{\sqrt{\alpha'}}{R}n^m+\frac{R}{\sqrt{\alpha'}}w^m,
\end{align}
\end{subequations}
where R is the scale factor in the compact dimension m, $n^m$ is  an integer(KK momentum charge),and $w^m$ is the winding number(how many times the string winds around this toroidal direction),then we have two kinds of modes:the oscillating mode and the winding mode.\\
This solution(so the action) possesses an additional symmetry, for the mass spectrum of these strings called the toroidal duality (T-duality), the mass spectrum is invariant under 
$$\frac{\sqrt{\alpha'}}{R} \leftrightarrow \frac{R}{\sqrt{\alpha'}}$$
$$n^m \leftrightarrow w^m$$
This implies that the physics in the case of very large scale factors is the same as in very small scale factors.\\
During this period of compactified universe its contents were one dimensional strings only, some literature extended it to include D branes$[83]$, then strings' world sheets begin to intersect and opposite winding modes annihilate, and as the energy of the winding modes is proportional to the scale factor(or the radius of the compact dimension) when it annihilates it takes less energy to expand so the dimension expands, and as one dimensional strings' world sheets can span a 3D space it is much more likely that they intersect in a 3D space and above that the intersection probability is zero, this means that only three dimensions will expand. In the case of higher dimensional D-brane added it was shown that these conclusions do not change.\\
This model also eliminates the big bang singularity completely due to the existence of a maximum possible temperature(the Hagedorn temperature), to see this consider a box of strings at thermal equilibrium, if the size of the box decreases its temperature will increase as usual matter,if the box continue shrinking at some point the energy of the momentum modes begins to transfer to the oscillation modes this prevents any rise of temperature, shrinking further to the string scale the energy is transferred to the winding mode leading to temperature drop so in the whole process of shrinking the temperature is bounded by the Hagedorn temperature as in fig[6] i.e. whatever the size of the box (the universe) there is no singularities.\\  
\begin{figure}[hbtp]
\centering
\includegraphics[scale=0.3]{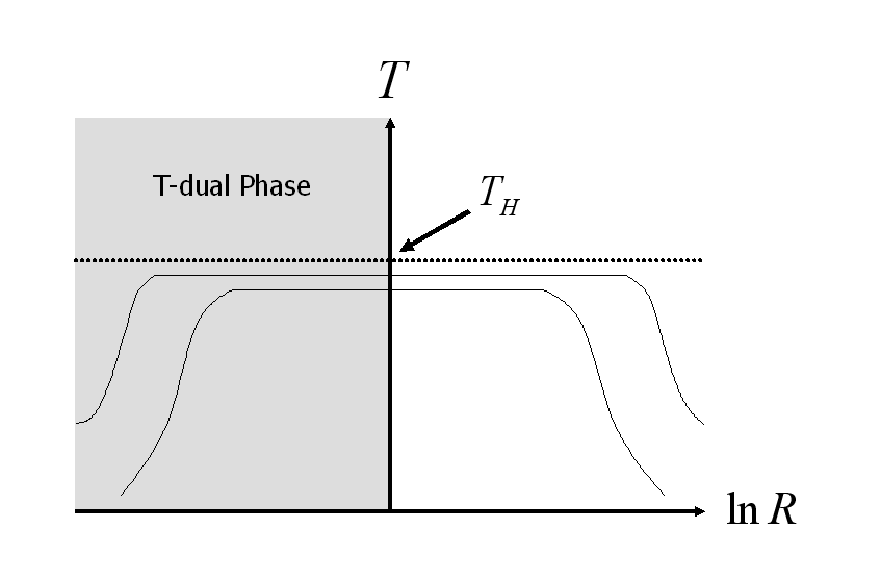}
\caption{As the radius R decreases the temperature firstly increases then stay constant as it crosses to the T-dual phase and decreases again due to the energy transmission to the winding modes$[94]$}
\end{figure}

The next topic to be discussed is moduli stabilization, string gas cosmology provides a mechanism form the theory itself to stabilize geometric moduli (the radions)$[84]$. However, the dilaton requires an additional ingredient to be inserted to be stabilized$[85]$ as we will see.\\

\textbf{1-Geometric moduli}.\\

This type of moduli is stabilized by the presence of enhanced symmetry states coming from the existence of momentum and winding modes, the momentum modes prevents the dimensions from contraction as its energy is proportional to 1/R, the winding modes prevents expansion as disused earlier, then the preferred size of the dimensions is the self dual value $R=1$, at this value the modes are massless and are seen as radiation from the 3D point of view solving the radiation abundance problem, in case of bosonic strings it is not the lowest energy stated due to the presence of tachyons, but in superstring theory it is because tachyons are ruled out by the GSO projection. Many literature disused geometric moduli stabilization, and even extend the mechanism to more general systems. There is an attempt to use dilaton gravity to do the job but dilaton is rapidly changing in the Hagedorn phase so the frame will not be static as wanted, so we consider the coupling of the string gas to Einstein gravity instead, this gives the metric
\begin{equation}
ds^2=dt^2-a(t)^2dx^2-\sum_{\alpha=1}^6b_{\alpha(t)}^2dy_{\alpha}^2 ,
\end{equation}
where x are the large dimensions coordinates and y are the small dimensions coordinates.\\
This gives the equations for the radions $b_{\alpha}$
\begin{equation}
\ddot{b_{\alpha}}+(3H+\sum_{\beta=1,\beta\neq \alpha}^6 \frac{\dot{b_{\alpha}}}{b_{\alpha}})\dot{b_{\alpha}}=\sum_{n,m}8 \pi G_N \frac{\mu_{m,n}}{\epsilon_{m,n}}S
\end{equation}
where $\mu_{m,n}$ is the number density of states with momentum mode n and winding mode m and $\epsilon_{m,n}$ is the energy of the state and S is a source term deduced from the mass spectrum formula and the level matching condition to be
\begin{equation}
S=\sum_{\alpha}(\frac{m_{\alpha}}{b_{\alpha}})^2-\sum_{\alpha}n_{\alpha}^2b_{\alpha}^2+\frac{2}{D-1}[n.n+n.m+2[N-2]]
\end{equation}
where N is the number of right moving oscillator states and m,n are the vectors representing the momentum and winding modes in the extra (small) dimensions.\\
The first two terms of S in eq.(143) represent an effective potential with a stable minimum at the self dual radius $b_{\alpha}=1$, then for the moduli to be stabilized the third term must vanish to prevent any positive potential contribution, this is true if and only if the states are masseless.\\

\textbf{2-Dilaton stabilization}.\\

There are many mechanisms proposed to stabilize the dilaton because it requires an additional ingredient, the most promising mechanism is the gaugino condensation. This mechanism modifies the superpotential of the theory as
\begin{equation}
W \rightarrow W-Ae^{-1/g_s}
\end{equation}
where A is a constant and $g_s$ is the string coupling constant.\\
The potential s derived from
\begin{equation}
V=\frac{e^{\mathcal{K}}}{M_p^2}[\mathcal{K}^{AB}D_A W D_{\bar{B}}\bar{W}-3|W|^2]
\end{equation}
where the derivative operator is defined as
\begin{equation}
D_A=\partial_A+[\partial_A\mathcal{K}]
\end{equation}
where $\mathcal{K}$ is the Kaehler potential and A,B in eq.(145) run over moduli fields. Since the superpotential is independent of the geometric moduli, we change the variables to 
$$S=e^{-\Phi}+ia$$
$$\Phi=2\phi -6lnb$$
a is called axion.\\

The potential given at eq.(145) is then reduced to
\begin{equation}
V=\frac{e^{\mathcal{K}}}{M_p^2}\mathcal{K}^{ab}D_A W D_{\bar{B}}\bar{W}
\end{equation}
lifting to 10 dimensions and expanding about the minimum say $\Phi_0$ we get
\begin{equation}
V=\frac{M_{10}^{16}V'}{4}e^{-\Phi_0}a_0^2A^2[a_0-\frac{3e^{\Phi_0}}{2}]^2e^{-2a_0e^{-\Phi_0}}e^{-3\phi/2}[b^6e^{-2\phi}-e^{-\Phi_0}]^2
\end{equation}
where $M_{10}$ is the Planck mass in 10D, V' is the volume of the compact space and $a_0$ is a constant.\\
If we expand the equations of motion about the self dual radius and the value of the dilaton minimizing V, it is proves that it is a stable minimum, this means that the addition of the potential stabilizes the dilaton without destabilize the radion i.e.in string gas cosmology all the moduli are fixed simultaneously.\\

\textbf{- Structure formation and the horizon problem.}\\

In string gas cosmology cosmological perturbations which lead to structure formation is due to thermal fluctuations of the strings$[87,88]$, this produces a nearly scale invariant spectrum for both curvature perturbations with the usual red tilt and gravitational waves but with a blue tilt in contrast with inflation$[89]$.\\
In the same context the horizon problem is solved due to the long Hagedorn phase which buys time for the horizon to be much larger than the Hubble radius, the perturbation modes' wavelengths remain constant during the Hagedorn phase while the Hubble's radius decreases quickly such that at the end of the Hagedorn phase the Hubble radius is larger than the perturbations' wavelengths, then Hubble radius increases again so the perturbations re-enter the horizon, this solves the horizon problem in a nearly similar way to inflation, moreover the fluctuation modes' wavelength observed is of the order of 1mm which is much larger than Planck's length so string gas cosmology evades the trans-Planckian problem$[89]$.(see also $[86]$ for more detailed review.)\\

\textbf{- The flatness and unwanted relics problems}.\\

In a recent paper by Vahid Kamali and Robert Brandenberger$[91]$ they proposed a solution for the flatness problem by combining string gas cosmology with power law inflation, the idea is adding a phase of inflation between the Hagedorn phase and the radiation dominated era, this solves the flatness problem as in the classical inflation. This problem is evaded if we consider the Hagedorn phase is a part of a crunch similar to the brane world cosmologies.\\

On the other hand the unwanted relics problem is not solved completely, a paper by Diana Battefeld and Thorsten Battefeld$[92]$ suggests two possible regimes to solve this problem, the first one is to incorporate a phase of reheating such that the reheating temperature is less than the required to  produce the relics, although this method succeeded to dilute gravitinos, it failed with magnetic monopoles. The second regime is to add an inflation phase after the Hagedorn phase similar to the solution to the flatness problem.\\
The problem of adding a phase of inflation is that if it lasts more than 12e-folds the dilution breaks the string gas approximation so few cosmic strings remains this means that the theory does not anymore, another problem is the removal of moduli stabilization mechanism by the inflation. The moduli stabilization mechanism used is the quantum trapping mechanism$[100]$, inflation removes the stabilization mechanism openning a new problem for this model.\\

\textbf{-Summary}:
According to string cosmology the universe began as a small compactified space filled with string gas(and maybe D-branes), and all dimensions are assumed to be toroidal, the string gas are in thermal equilibrium with temperature near Hagedorn temperature, strings have winding modes on collision these modes annihilate leading to the expansion of three dimensions only as one dimensional strings can span a 3D space only, the thermal fluctuations in the string gas are the seeds of the structure formation after the end of the rediation dominated era, this model solves the horizon, trans-Plankian and singularity problems, but requires a combination with power law inflation to solve the flatness problem, however it does not solve the unwanted relics problem completely.\\

\textbf{Open Problems}:\\

1- The expansion of only three dimensions is derived from the annihilation of winding modes by the collision of the strings, however it is a classical view in fact there are long range forces(exchange of closed strings) by which strings can interact and annihilate, so the probability for expansion of more than three dimensions is not zero.\\
2- Sting cosmology uses the dilaton gravity as a background instead of Einstein's gravity because Einstein's gravity does not respect T-duality, however in dilaton gravity the dilaton evolves rapidly with time in the Hagedorn phase so the frame will not be static, so the background of the theory is yet unknown.\\
3- we have no mathematical model for the hagedorn phase nor any quantitative description of it.\\
4- There is a hierarchy problem in the model, if the string scale is of the order of $10^{17}$ Gev (the preferred energy scale in heterotic superstring theory), the size of the universe during the Hagedorn phase will be much larger than the string scale which is unnatural to assume.\\
This problem is solved in models with a contraction phase preceeds the Hagedorn phase, however this introduces a new problem: How could we maintain a long enough Hagedorn phase to reach thermal equilibrium over all the scales?\\
Finally, we note that string Gas Cosmology was reviewed and its challenges presented in$[93-97]$

\subsection{\textbf{Other models}}
There are some models which use string theory in cosmology to solve certain problems not to provide a description of the early universe like Randall-Sundrum first and second models$[98,99]$, the models attempt to solve the hierarchy problem in the standard model by assuming that the world is a 5D bulk bounded by two branes, the first brane is a brane where gravity is strong (called the Planck brane or the gravity brane), and the other where our universe is printed and Gravity is weak(called Tev brane or the weak brane), the second model RS2 uses one brane in an infinite 5D bulk(assuming that the other brane is infinitely far away), these models suffer from two fine tuning problems: the bulk cosmological constant and the tension of the branes. The RS2 model was generalized to the use of a thick brane with different values of the cosmological constant on each side.\\
Some other models are proposed to calculate certain quantities like the model proposed by M.Vijaya,V.U.M. Rao and Y.Aditya$[97]$ to derive the f(R) gravity field equations in the presence of a bulk viscous fluid and the model by Christos Charmousis and Jean-Francois Dufaux$[102]$, they proposed a 5D spacetime with constant spatial curvature with a bulk cosmological constant and Gauss-Bonnet term, and found the general solution in addition to many models of viscous bulks and attempts to study Bianchi type models by string cosmology$[103-107]$.\\ 
Other models try to calculate the behaviour near the singularity like the holographic cosmology.\\
The Holographic principle stated that the physics of a space is encoded on the lower dimensional boundary, this was applied in string theory by Susskind$[108]$,but was applied in cosmology by Maldacena$[109]$, and later was applied for 4D inflationary universe$[110]$, this model is promising as a cosmological model and fits the recent experimental data$[111]$, However, it is still a new model which needs a lot of work to be complete. \\
\subsection{\textbf{Outlook}}
A possible solution to the particle-antiparticle asymmetry can be derived from the new ekpyrotic model, after the collision particles are produced on the visible brane and antiparticles on the hidden brane as a back reaction to the collision, this preserves the symmetry of particles and antiparticles in the system of the two branes. This separation may be caused by some difference in graviton-visible brane interaction and graviton-hidden brane interaction due to a prior interaction with a viscous fluid filling the bulk between the two branes.\\
This kind of interaction with the bulk fluid can flip the gravitons coming from the hidden brane so that we see it as antigravity. Similar ideas were proposed by Abdel-Raouf $[112]$ where after the big bang particles were driven by gravity to form our universe and antiparticles were driven by antigravity forming an antiuniverse. This creates a universe-antiuniverse system which possess plausible physical properties, for example this system is CTP symmetric unlike the creation of a system of one universe $[113]$. Furthermore, it is suggested that the universe and antiuniverse may have a stable overlap region in which matter, and antimatter may form exotic molecular structures accompanied with continues creation and annihilation of particle-antiparticle pairs. The traces of antiparticles occurring in our universe are the ones escaped from this process $[114-116]$.\\

  \newpage

\section{ \textbf{References}}
$[1]$ R. M. Wald, General Relativity. University of Chicago Press, 1984.  doi:10.7208/chicago/9780226870373.001.0001.\\
$[2]$ R.P. Kirshner,  A.J.Oemler, P.L. Schechter, S.A. Shectman, A million cubic megaparsec void in Bootes. Astrophys. J. 248, L57 (1981).\\
$[3]$ A. Friedmann,Z. Phys. 10, 377(1922);ibid. 21, 326(1924).\\
$[4]$ H.P. Robertson, Astrophys. J. 82, 284(1935);ibid.,83,187,257(1936).\\
$[5]$ A.G. Walker, On Milne's Theory of World‐Structure, Proc. Lond. Math. Soc. (2)42,90(1936)\\
$[6]$ A. Friedmann,Z. Phys. 16, 377(1922);ibid. 21, 326(1924).\\
$[7]$ R.H. Dicke, Gravitation and the Universe. Amer Philosophical Society, 1970.\\
$[8]$ A. Lightman, Ancient light:Our Challenging View of the Universe. Harvard University Press, 1993.\\
$[9]$ R. Carrigan, W.P. Trower, Magnetic Monopoles. (Springer US, 1983). doi:10.1007/978-1-4615-7370-8.\\
$[10]$ T.W.B. Kibble, Topology of cosmic domains and strings. J. Phys. A. Math. Gen. 9, 1387–1398 (1976).\\
$[11]$ T.W.B. Kibble, Some implications of a cosmological phase transition. Phys. Rep. 67, 183–199 (1980).\\
$[12]$  W.M. Yao et al, Review of Particle Physics. J. Phys. G Nucl. Part. Phys. 33, 1–1232 (2006).\\
$[13]$ A.A. Starobinsky, A new type of isotropic cosmological models without singularity. Phys. Lett. B 91, 99–102 (1980).\\
$[14]$ A.H. Guth, "The inflationary universe universe: A possible solution to the horizon and the flatness" Phys. Rev. D 23(2),347-356(1981).\\
$[15]$ C. Smeenk, False Vacuum: Early Universe Cosmology and the Development of Inflation. Universe Gen. Relativ. 223–257 (2005), doi.org/10.1007/0-8176-4454-7$_{-}$13\\
$[16]$ A. Linde, Hybrid inflation. Phys. Rev. D 49, 748–754 (1994).\\
$[17]$ C. Armendáriz-Picón, T. Damour, V. Mukhanov,  k-Inflation. Phys. Lett. B 458, 209 – 218 (1999).\\
$[18]$ A. Maleknejad, M.M. Sheikh-Jabbari,  Gauge-flation: Inflation from non-Abelian gauge fields. Phys. Lett. B 723, 224 – 228 (2013).\\
$[19]$ F. Lucchin, S. Matarrese,  Power-law inflation. Phys. Rev. D 32, 1316–1322 (1985).\\
$[20]$ A.D. Linde, Chaotic inflation. Phys. Lett. B 129, 177–181 (1983).\\
$[21]$ M. Berg,E. Pajer, S. Sjors,  Dante’s Inferno. Eos, Trans. Am. Geophys. Union 75, 378 (2009).\\
$[22]$ A.D. Linde,  A new inflationary universe scenario: A possible solution of the horizon, flatness, homogeneity, isotropy and primordial monopole problems. Phys. Lett. B 108, 389–393 (1982).\\
$[23]$ A. Albrecht, P.J. Steinhardt, Cosmology for Grand Unified Theories with Radiatively Induced Symmetry Breaking. Phys. Rev. Lett. 48, 1220–1223 (1982).\\
$[24]$ S. Lahiri,  Dirac Born-Infeld inflation under constant roll conditions. (2018).\\
$[25]$ A.D. Sakharov, Violation of CP in variance, C asymmetry, and baryon asymmetry of the universe. Sov. Phys. Uspekhi 34, 392–393 (1991).\\
$[26]$ J. Martin, R.H. Brandenberger,  Trans-planckian problem of inflationary cosmology. Phys. Rev. D 63, (2001).\\
$[27]$ H. Năstase, Cosmology and String Theory. vol. 197, Springer International Publishing, 2019.\\
$[28]$ R.J. Adler, B. Casey, O.C. Jacob, Vacuum catastrophe: An elementary exposition of the cosmological constant problem. Am. J. Phys. 63, 620–626 (1995).\\
$[29]$ M.B. Green,J.H. Schwartz, E. Witten, Superstring theory. Vol. 1: Introduction, . Cambridge, UK: Univ. Pr. (1987) 469 P.(Cambridge monographs in mathematical physics).\\
$[30]$ L. Susskind, Dual symmetry theory of hadrons. 1, Nuovo Cim. A69(1970) 457-496.\\
$[31]$ Y. Nambu, Dual model of hadrons, . EFI-70-07.\\
$[32]$ Y. Nambu,  (1970). "Quark model and the factorization of the Veneziano amplitude." In R. Chand (ed.), Symmetries and quark models (pp. 269–277).\\
$[33]$ S. Mukhi, "String theory: a perspective over the last 25 years"[arXiv:11 10.2569v3].\\
$[34]$ P. Ramond, Dual Theory for Free Fermions. Phys. Rev. D 3, 2415–2418 (1971).\\
$[35]$ A. Neveu, J.H. Schwarz, Tachyon-free dual model with a positive-intercept trajectory. Phys. Lett. B 34, 517–518 (1971).\\
$[36]$ J. Wess, B. Zumino,  Supergauge transformations in four dimensions. Nucl. Phys. B 70, 39–50 (1974).\\
$[37]$ J.H. Schwarz, String Theory Origins of Supersymmetry. 0–13 (2000) doi:10.1016/S0920-5632(01)01492-X.\\
$[38]$ F. Gliozzi, J. Scherk, D. Olive,  Supersymmetry, supergravity theories and the dual spinor model. Nucl. Phys. B 122, 253–290 (1977).\\
$[39]$ M.B. Green, J.H. Schwarz, Anomaly cancellations in supersymmetric D = 10 gauge theory and superstring theory. Phys. Lett. B 149, 117–122 (1984).\\
$[40]$ C. Montonen, D. Olive,  Magnetic monopoles as gauge particles? Phys. Lett. B 72, 117–120 (1977).\\
$[41]$ H. Osborn, Topological charges for N = 4 supersymmetric gauge theories and monopoles of spin 1. Phys. Lett. B 83, 321–326 (1979).\\
$[42]$ A. SEN, Strong–Weak Coupling Duality in Four-Dimensional String Theory. Int. J. Mod. Phys. A 09, 3707–3750 (1994).\\
$[43]$ C. Vafa, E. Witten,  A strong coupling test of S-duality. Nucl. Phys. B 431, 3–77 (1994).\\
$[44]$ M.B. Green, J.H. Schwarz,  Anomaly cancellations in supersymmetric D = 10 gauge theory and superstring theory. Phys. Lett. B 149, 117–122 (1984).\\
$[45]$ K. Kikkawa, M. Yamasaki,  Casimir effects in superstring theories. Phys. Lett. B 149, 357–360 (1984).\\
$[46]$ A. Strominger, Superstrings with torsion. Nucl. Phys. B 274, 253–284 (1986).\\
$[47]$ B. de Wit, J. Hoppe, H. Nicolai,  On the quantum mechanics of supermembranes. Nucl. Phys. B 305, 545–581 (1988).\\
$[48]$ J. Polchinski, Dirichlet Branes and Ramond-Ramond Charges. Phys. Rev. Lett. 75, 4724–4727 (1995).\\
$[49]$ A. Strominger, C. Vafa,  Microscopic origin of the Bekenstein-Hawking entropy. Phys. Lett. Sect. B Nucl. Elem. Part. High-Energy Phys. 379, 99–104 (1996).\\
$[50]$ J. Maldacena, The large-N limit of superconformal field theories and supergravity. Int. J. Theor. Phys. 38, 1113–1133 (1999).\\
$[51]$ T. Banks, W. Fischler, S.H. Shenker, L. Susskind,  M theory as a matrix model: A conjecture. Phys. Rev. D 55, 5112–5128 (1997).\\
$[52]$ J.M. Maldacena, The Large N Limit of Superconformal Field Theories and Supergravity. (1997) doi:10.1023/A:1026654312961.\\
$[53]$ E. Witten, Anti de sitter space and holography. Adv. Theor. Math. Phys. 2, 253–290 (1998).\\
$[54]$ S.S. Gubser, I.R. Klebanov, A.M. Polyakov,  Gauge theory correlators from non-critical string theory. Phys. Lett. Sect. B Nucl. Elem. Part. High-Energy Phys. 428, 105–114 (1998).\\
$[55]$ A. Jaffe, F. Quinn,  "Theoretical mathematics": Toward a cultural \\synthesis of mathematics and theoretical physics. Bull. Am. Math. Soc. 29, 1–14 (1993).\\
$[56]$ C. Rovelli,  A critical look at strings. (2011) doi:10.1007/s10701-011-9599-3.\\
$[57]$ M.B. Green, J.H. Schwarz, E. Witten,  Superstring Theory. (Cambridge University Press, 2012). doi:10.1017/CBO9781139248570.\\
$[58]$ M.B. Green, J.H. Schwarz, E. Witten,  Superstring Theory. (Cambridge University Press, 2012). doi:10.1017/CBO9781139248563.\\
$[59]$ M.B. Green, J.H. Schwarz, E. Witten,  Superstring Theory. (Cambridge University Press, 2012). doi:10.1017/CBO9781139248563.\\
$[60]$ J. Polchinski, String Theory. (Cambridge University Press, 1998).\\ doi:10.1017/CBO9780511618123.\\
$[61]$ B. Zwiebach, A First Course in String Theory. (Cambridge University Press, 2009). doi:10.1017/CBO9780511841620.\\
$[62]$ H. Năstase, Cosmology and String Theory. vol. 197 (Springer International Publishing, 2019).\\
$[63]$ G. Dvali, S.H. Henry Tye,"Brane inflation," Phys.Lett. B450(1999) 72-82[arXiv.hep-ph/9812483v1].\\
$[64]$ P. Hořava, E. Witten, Heterotic and type I string dynamics from eleven dimensions. Nucl. Phys. B 460, 506–524 (1996).\\
$[65]$ P. Hořava, E. Witten, Eleven-dimensional supergravity on a manifold with boundary. Nucl. Phys. B 475, 94–114 (1996).\\
$[66]$ P. Binétruy, C. Deffayet, D. Langlois,  Non-conventional cosmology from a brane universe. Nucl. Phys. B 565, 269–287 (2000).\\
$[67]$ C.P. Burgass, M. Majumdar, D. Nolte,F. Quevedo, G. Rajesh, R.J. Zhang,"the inflationary brane anti-brane universe," JHEP 1202, 040(2012) [arXiv:hep-th/0105204].\\
$[68]$A. Sen, Stable non-BPS bound states of BPS D-branes, J. High Energy Phys. 9808(1998)010[hep-th/9805016].\\
A. Sen, Tachyon condensation on the brane-antibrane system, J. High Energy Phys. 9808(1998)012[hep-th/9805170].\\
$[69]$R. Brandenberger, C. Vafa, Superstrings in Te early universe, Nucl. Phys. B 316(1989)391;\\
$[70]$S. Alexander, R. Brandenberger, D. Easson, Brane gases in the early universe, Phys. Rev D62(2000) 103509[hep-th/0005212].\\
$[71]$ F. Quevedo, Lectures on string/brane cosmology. Class. Quantum Gravity 19, 5721–5779 (2002).\\
$[72]$ N. Turok, Ekpyrotic universe: Colliding branes and the origin of the hot big bang. Phys. Rev. D - Part. Fields, Gravit. Cosmol. 64, 24 (2001).\\
$[73]$ J. Khoury, B.A. Ovrut, N. Seiberg, P.J. Steinhardt,  N. Turok,  From big crunch to big bang. Phys. Rev. D - Part. Fields, Gravit. Cosmol. 65, 8 (2002).\\
$[74]$ N. Turok, Cosmic evolution in a cyclic universe. Phys. Rev. D - Part. Fields, Gravit. Cosmol. 65, 20 (2002).\\
$[75]$Brandenberger, R., Heisenberg, L. \& Robnik, J. Non-singular black holes with a zero-shear S-brane. J. High Energy Phys. 2021, 90 (2021).\\
$[76]$ Gutperle, M. \& Strominger, A. Spacelike branes. J. High Energy Phys. 6, 395–413 (2002).\\
$[77]$Brandenberger, R. \& Wang, Z. Ekpyrotic cosmology with a zero-shear S-brane. Phys. Rev. D 102, 23516 (2020).\\
$[78]$ Brandenberger, R., Dasgupta, K. \& Wang, Z. Reheating after S-brane ekpyrosis. Phys. Rev. D 102, 63514 (2020).\\
$[79]$ E.I. Buchbinder, J. Khoury, B.A. Ovrut,  New ekpyrotic cosmology. Phys. Rev. D - Part. Fields, Gravit. Cosmol. 76, (2007).\\
$[80]$ L. McAllister, E. Silverstein,  String cosmology: A review. Gen. Relativ. Gravit. 40, 565–605 (2008).\\
$[81]$ A. Linde,  Inflationary Theory versus Ekpyrotic/Cyclic Scenario. (2002).\\
$[82]$ A. Nayeri, R.H. Brandenberger,  C. Vafa,  Producing a Scale-Invariant Spectrum of Perturbations in a Hagedorn Phase of String Cosmology. 1–4 (2005) doi:10.1103/PhysRevLett.97.021302.\\
$[83]$ S. Alexander, R. Brandenberger, D. Easson,  Brane gases in the early Universe. Phys. Rev. D - Part. Fields, Gravit. Cosmol. 62, 7 (2000).\\
$[84]$ S.P. Patil, R.H. Brandenberger, “The cosmology of massless string modes,” JCAP 0601, 005 (2006) [arXiv:hep-th/0502069].\\
$[85]$ R.J. Danos, A.R. Frey, R.H. Brandenberger, “Stabilizing moduli with thermal matter and nonperturbative effects,” Phys. Rev. D 77, 126009 (2008) [arXiv:0802.1557 [hep-th]].\\
$[86]$ N. Arkani-Hamed, P. Creminelli, S. Mukohyama,  M. Zaldarriaga,  Ghost inflation. J. Cosmol. Astropart. Phys. 1–18 (2004) doi:10.1088/1475-7516/2004/04/001.\\
$[87]$A. Nayeri, R.H. Brandenberger, C. Vafa, “Producing a scale-invariant spectrum of perturbations in a Hagedorn phase of string cosmology,” Phys. Rev. Lett. 97, 021302 (2006) [arXiv:hep-th/0511140].\\
$[88]$ R.H. Brandenberger, A. Nayeri, S.P. Patil, C. Vafa, “String gas cosmology and structure formation,” Int. J. Mod. Phys. A 22, 3621 (2007) [hep-th/0608121].\\
$[89]$ R.H. Brandenberger, A. Nayeri, S.P. Patil, C. Vafa, “Tensor Modes from a Primordial Hagedorn Phase of String Cosmology,” Phys. Rev. Lett. 98, 231302 (2007) [hep-th/0604126].\\
$[90]$ J.Erdmenger, String Cosmology, Wiley-VCH, Weinheim, 2009.\\
$[91]$ V. Kamali, R. Brandenberger,  Creating spatial flatness by combining string gas cosmology and power law inflation. Phys. Rev. D 101, 1–9 (2020).\\
$[92]$ D. Battefeld, T. Battefeld,  The Relic Problem of String Gas Cosmology. (2009) doi:10.1103/PhysRevD.80.063518.\\
$[93]$ R. Brandenberger,  (2009). String Gas Cosmology.\\ 10.1002/9783527628063.ch6. \\
$[94]$ T. Battefeld, S. Watson, String gas cosmology. Rev. Mod. Phys. 78, 435–454 (2006).\\
$[95]$ R.H. Brandenberger, Challenges for String Gas Cosmology. 1–10 (2005).\\
$[96]$ R.H. Brandenberger,  String gas cosmology after Planck. Class. Quantum Gravity 32, (2015).\\
$[97]$ R.H. Brandenberger,  String gas cosmology: Progress and problems. Class. Quantum Gravity 28, (2011).\\
$[98]$ L. Randall, R. Sundrum,  Large mass hierarchy from a small extra dimension. Phys. Rev. Lett. 83, 3370–3373 (1999).\\
$[99]$ L. Randall,  R. Sundrum,  An alternative to compactification. Phys. Rev. Lett. 83, 4690–4693 (1999).\\
$[100]$ L. Kofman, A. Linde, X. Liu, A. Maloney, L. McAllister, E. Silverstein. Beauty is attractive: Moduli trapping at enhanced symmetry points. J. High Energy Phys. 8, 671–711 (2004).\\
$[101]$ M.V. Santhi, V.U.M. Rao, Y. Aditya,  Bulk viscous string cosmological models in f ( R ) gravity. Can. J. Phys. 96, 55–61 (2018).\\
$[102]$ C. Charmousis, J.F. Dufaux,  General Gauss-Bonnet brane cosmology. 1–21 (2002) doi:10.1088/0264-9381/19/18/304.\\
$[103]$ A. Pradhan, D.S. Chouhan,  Anisotropic Bianchi type-I models in string cosmology. Astrophys. Space Sci. 331, 697–704 (2011).\\
$[104]$ A. Banerjee, A.K. Sanyal, S. Chakraborty,  String cosmology in Bianchi I space-time. Pramana 34, 1–11 (1990).\\
$[105]$ S.R. Roy, S.K. Banerjee, Bianchi type II string cosmological models in general relativity. Class. Quantum Gravity 12, 1943–1948 (1995).\\
$[106]$ W. Xing-Xiang,  Bianchi Type-III String Cosmological Model with Bulk Viscosity in General Relativity. Chinese Phys. Lett. 22, 29–32 (2005).\\
$[107]$ A.K. Yadav, Bianchi-V string cosmology with power law expansion in f (R, T) gravity. Eur. Phys. J. Plus 129, (2014).\\
$[108]$ L. Susskind,  The world as a hologram. J. Math. Phys. 36, 6377–6396 (1995).\\
$[109]$ J. Maldacena,  Non-gaussian features of primordial fluctuations in single field inflationary models. J. High Energy Phys. 7, 233–264 (2003).\\
$[110]$ P. McFadden, K. Skenderis,  Holography for cosmology. Phys. Rev. D - Part. Fields, Gravit. Cosmol. 81, (2010).\\
$[111]$ N. Afshordi, C.  Corianò,  L.D. Rose, E. Gould, K. Skenderis,  From Planck Data to Planck Era: Observational Tests of Holographic Cosmology. Phys. Rev. Lett. 118, 1–8 (2017).\\
$[112]$ Abdel-Raouf, Mohamed \& Raouf, Abdel. (2013). Possible Existence of Overlapping Universe and Antiuniverse.\\
$[113]$ L. Boyle, K. Finn, N. Turok,  CPT-Symmetric Universe. Phys. Rev. Lett. 121, 251301 (2018).\\
$[114]$ Abdel-Raouf Mohamed (2011) Possible Formation of Overlapping Universe and Antiuniverse, ID:1, DPF 2011, held at Brown University, August 8-13: \\
$[115]$ Abdel-Raouf, Mohamed. (2019). Matter-Antimatter Physics at Low Energy. Journal of Physics: Conference Series. 1253. 012006. 10.1088/1742-6596/1253/1/012006. \\
$[116]$ Abdel-Raouf, M.A.
(2020) Novel Consequences of Coexistence
of Matter and Antimatter in Nature. Journal
of High Energy Physics, Gravitation and
Cosmology, 6, 251-258.
https://doi.org/10.4236/jhepgc.2020.62019\\

\end{document}